\documentclass[10pt, twocolumn,twoside,pre]{revtex4-1}
\usepackage{color}
\usepackage{siunitx}
\usepackage{mathtools}
\usepackage[normalem]{ulem}
\usepackage{color}

\begin{document}
\title{Localization-delocalization transition for light particles in turbulence}

\author{Ziqi Wang}
\affiliation{Department of Applied Physics and Science Education, Eindhoven University of Technology, 5600 MB Eindhoven, Netherlands}			
\author{Xander M. de Wit} 
\affiliation{Department of Applied Physics and Science Education, Eindhoven University of Technology, 5600 MB Eindhoven, Netherlands}			
\author{Federico Toschi}
\affiliation{Department of Applied Physics and Science Education, Eindhoven University of Technology, 5600 MB Eindhoven, Netherlands; CNR-IAC, I-00185 Rome, Italy}
\email{F.Toschi@tue.nl}

\begin{abstract}
{Small bubbles in fluids rise to the surface due to Archimede’s force. Remarkably, in turbulent flows this process is severely hindered by the presence of vortex filaments, which act as moving potential wells, dynamically trapping light particles and bubbles. Quantifying the statistical weights and roles of vortex filaments in turbulence is, however, still an outstanding experimental and computational challenge due to their small scale, fast chaotic motion, and transient nature. 
	Here we show that, under the influence of a modulated oscillatory forcing, the collective bubble behaviour switches from a dynamically localized to a delocalized state. Additionally, we find that by varying the forcing frequency and amplitude, a remarkable resonant phenomenon between light particles and small-scale vortex filaments emerges, likening particle behavior to a forced damped oscillator. We discuss how these externally actuated bubbles can be used as a type of microscopic probe to investigate the space-time statistical properties of the smallest turbulence scales, allowing to quantitatively measure physical characteristics of vortex filaments. We develop a superposition model that is in excellent agreement with the simulation data of the particle dynamics which reveals the fraction of localized/delocalized particles as well as characteristics of the potential landscape induced by vortices in turbulence.
	Our approach paves the way for innovative ways to accurately measure turbulent properties and to the possibility to control light particles and bubble motions in turbulence with potential applications to medical imaging, drug/gene delivery, chemical reactions, wastewater treatment, and industrial mixing.
}
\end{abstract}

%\dates{This manuscript was compiled on \today}
%\doi{\url{www.pnas.org/cgi/doi/10.1073/pnas.XXXXXXXXXX}}

%\begin{document}

\maketitle
%\thispagestyle{firststyle}
%\ifthenelse{\boolean{shortarticle}}{\ifthenelse{\boolean{singlecolumn}}{\abscontentformatted}{\abscontent}}{}
Turbulence, with its complex multiscale and multitime correlations \cite{frisch1995turbulence}, pervades nature and industrial processes, yet our current understanding and ability to manipulate it is still rather limited. A special role in turbulence is commonly attributed to the presence of coherent structures that, in homogeneous and isotropic turbulence (HIT), are represented by small-scale vortex filaments \cite{jimenez1993structure,jimenez1998characteristics, sreenivasan1997phenomenology, ghira2022characteristics, wallace2009twenty,johnson2024multiscale}. The extremely small scale, fast chaotic motion, and transient nature (finite lifetime \cite{biferale2010measurement}) of vortex filaments pose challenges and require huge efforts to capture, observe, and measure their dynamics, either experimentally or numerically, leaving their role in turbulence largely unexplored and somehow ambiguous \cite{buzzicotti2020statistical}. 
On the other hand, the core of vortex filaments can easily be visualized by simply introducing small bubbles in turbulent flows, as bubbles are dynamically attracted towards the core of vortex filaments (even as the vortex filaments swiftly sweep around in turbulent flows), and, consequently, the average bubble rising velocities are much slower compared to those in still fluids\cite{mazzitelli2003relevance, calzavarini2008dimensionality, mathai2020bubbly, toschi2009lagrangian}. 
Therefore, understanding the interplay and bridging the dynamics of light particles (including bubbles) and vortex filaments offers a unique opportunity to better characterize small-scale structures in turbulence, as well as in controlling the macroscopic average rise dynamics of bubbles in turbulence.

In many real-world applications, turbulent flows are laden with small light-particle, examples including oceanography (air entrapment induced by surface wave breaking profoundly impacting the ocean-atmosphere exchange of energy and matter which is crucial for climate processes and marine biological structure evolution \cite{deike2016air}) and marine biology (forming, disturbing, and transporting high-density patches of plankton organisms in turbulent oceanic flows \cite{peters2000effects}), as well as chemical engineering (air bubbles introduced into chemical reactors enhance mixing and reaction rates \cite{martin2008bubbling}) and petroleum industry (oil droplets in turbulent pipelines experience complex flow patterns during transportation \cite{ismail2015review}). 
In many of these practical applications, the motion of the submerged light particles is not only affected by the background turbulence, but is simultaneously subjected to different types of external forces. This forcing includes pressure fluctuations induced either by vibration/motion \textcolor{black}{of} the turbulent flows themselves (e.g., \textcolor{black}{shaking the container \cite{seropian2023effect}}) or by externally applied acoustic fields \textcolor{black}{\cite{crum1970motion}} (widely used for precise medical interventions \cite{seo2010microfluidic,preetam2022emergence} such as targeted drug delivery, medical imaging, and therapy, etc.) as well as magnetic or electric forces directly acting on particles (as functionalized with magnetic materials/nanoparticles \cite{solovev2009catalytic}) or charged light particles (due to the presence of injection or space charges \cite{beroual1992behavior,seo2010microfluidic}), both of which enable manipulating particle responses. Particle behaviors can also be adjusted by modifying particle inertia parameters (particles display localization and delocalization transitions in turbophoresis \cite{belan2014localization,belan2015phase}) and by adding a random force ﬂuctuating in both space and time on particles (an order-disorder transition is identified in a system of particles \cite{wilkinson2003path, deutsch1985aggregation}). 

Regardless of the rich phenomenology and plenty of applications, a fundamental systematic study of the dynamics of light particles and their collective, macroscopic responses in turbulence (e.g. average speed and vortex trapping times), when subject to a general external forcing, is currently lacking. Moreover, also a quantitative connection of the tight physical duality between light particle dynamics and the turbulence small-scale structures has remained elusive.

Here, we introduce a method for probing small-scale structures in turbulence by applying periodic external forces to light particles (either solid, liquid, or gaseous particles whose density is much smaller than that of the background fluid) within homogeneous isotropic turbulence. Through direct numerical simulations and theoretical modeling, we identify three distinct dynamical regimes—localization, transition, and delocalization—each with unique motion patterns and residence times within vortices. {\textcolor{black}{Localization refers to the situation where light particles preferentially concentrate inside vortex filaments, and delocalization refers to the case where particles escape vortices or experience only a minimal constraining effect due to the vortices.}} This localization is dynamic because of the finite lifetime of vortex filaments, resulting in particles being released, separating explosively, and then getting trapped again by other vortices. A remarkable resonant phenomenon emerges between particles and vortices as the forcing amplitude increases, allowing for the detection of the natural frequency of the vortex filaments. We propose a perspective likening particle behavior to a forced damped oscillator and develop a superposition model that is in excellent agreement with DNS results, revealing characteristics of the potential landscape induced by vortices.
Our study offers valuable insights for manipulating light particle behavior and opens the door towards harnessing turbulence for various applications.

\begin{figure}[!h]
	\centering
	\includegraphics[width=1\linewidth]{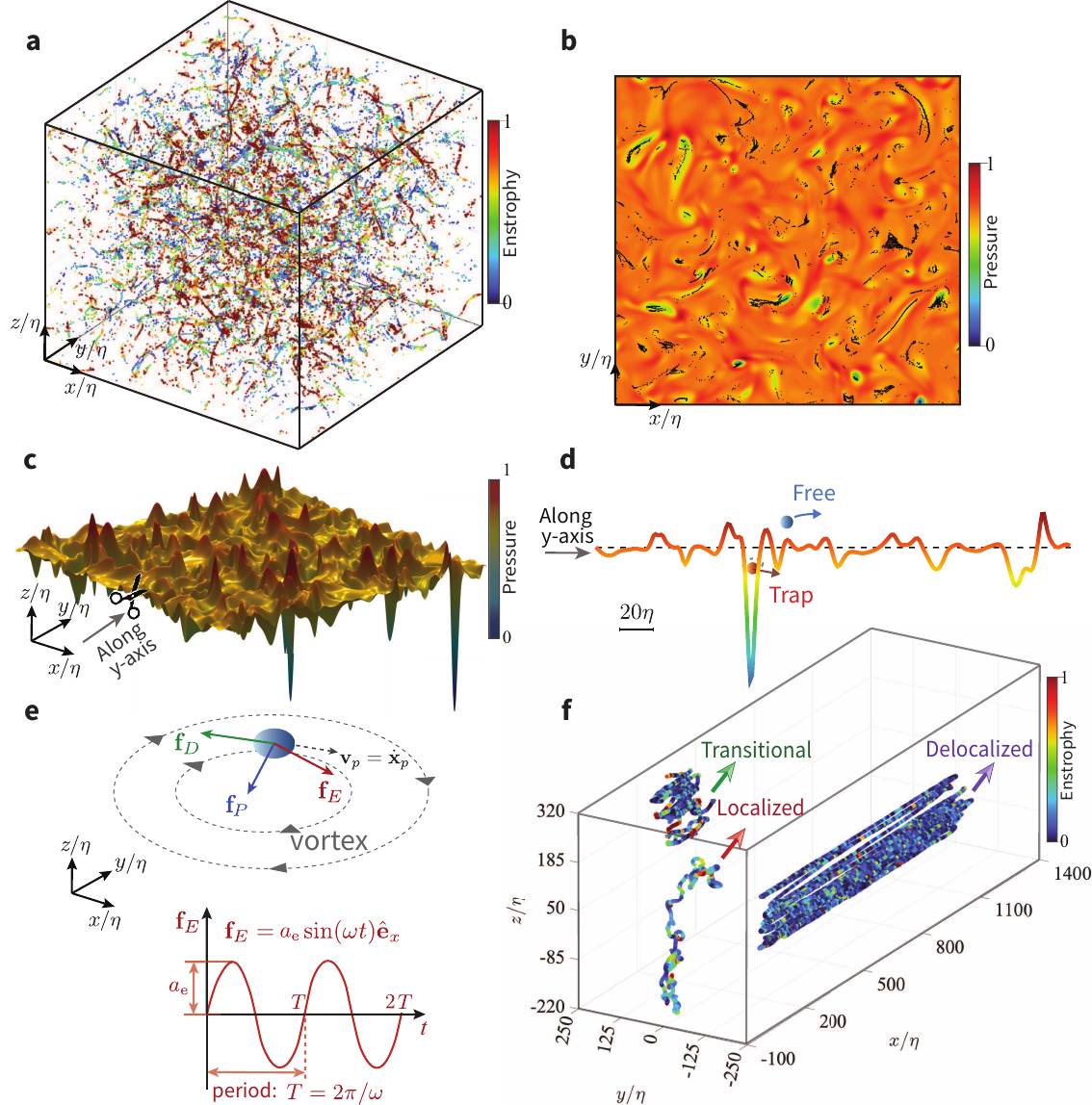}
	\vspace{-0.2cm}
	\caption{
		{\bf Potential landscape felt by light particles and its Lagrangian trajectories in homogeneous isotropic turbulence (HIT).} 
		{(a)} Snapshot of particle distributions. The results are from DNS of $N^3 = 256^3$ at St = 1.0, $\beta = 3$ (bubbles) with particle number $N_p =8.4\times10^6$.
		The particles are colored by the fluid enstrophy level at the particle position. It can be seen that light particles mostly concentrate preferentially in high-vorticity regions (represented by the red color), and filamentary shaped clusters are formed.
		(b) Instantaneous light particle positions and pressure level at $z=128$ corresponding to panel (a). Light particles are extracted within a slice with thickness ($\sim 3\eta$) whose planar center is at $z=128$.
		{(c)} Pressure landscape at the mid-height plane of {(a)} ($z=128$).  Cutting through the landscape along $y$ axis (at $x=128$) we have the pressure level as shown in {(d)}. One particle's instantaneous state, either trapped or free from the vortex is demonstrated in (d).
		{(c)} and {(d)} share the same colorbar. 
		{(e)} Forces felt by a light particle moving past a vortex. The forces due to drag ($\mathbf{f}_D$) and the net effect of added mass and pressure ($\mathbf{f}_P$) are represented by green-colored and blue-colored arrows respectively. A periodic external force oriented in the $x$ direction, $\mathbf{f}_E = a_{\text{e}} \sin(\omega t) \hat{\mathbf{e}}_x$, is applied on the particle.  
		{(f)} Lagrangian trajectories of light particles. 
		Lagrangian trajectories of light particles exhibit three distinct regimes based on the amplitude of the external force $a_{\text{e}}$ (while keeping fixed angular frequency $\omega$), i.e., localized (low $a_{\text{e}}$), transitional (intermediate $a_{\text{e}}$), and delocalized (high $a_{\text{e}}$). These regimes are distinguished by different particle oscillation amplitudes along the $x$ direction. To focus on the relative level of enstrophy and pressure, colormap values are normalized by the full range of enstrophy or pressure magnitude at the current time instant (see {(b)} and {(c)}) and along the estimated trajectories (see (f)).
	}
	\label{FIG1}
\end{figure}

\section*{Equation of motion for light particles subjected to a \\periodic external force in HIT}

To simplify the problem setting while retaining ingredients that capture the essential physics, we consider one of the simplest and yet nontrivial models of particles in HIT \cite{maxey1983equation,auton1988force,biferale2010measurement}, where passively advected, \textcolor{black}{one-way coupled} dilute suspensions of spherical light particles are considered, \textcolor{black}{which has been extensively validated through experiments and simulations \cite{toschi2009lagrangian, calzavarini2008dimensionality, biferale2010measurement}}. The particles experience forces due to drag ($\mathbf{f}_D$) as well as the net effect of added mass and pressure ($\mathbf{f}_P$). On the particles, we add a periodic external force of the form $\mathbf{f}_E = a_{\text{e}} \sin(\omega t) \hat{\mathbf{e}}_x$, where $a_{\text{e}}$ is the forcing amplitude and $\omega$ is the forcing frequency. The particles are light and one-way coupled point particles, whose acceleration is controlled by the Maxey–Riley equation of small particle ($\ll$ Kolmogorov length scale $\eta$) and no Fax\'en forces corrections \cite{maxey1983equation, mathai2020bubbly}. \textcolor{black}{This one-way coupling approach is valid in the dilute limit where the volume fraction of bubbles is small.}
The non-dimensionalized (by the Kolmogorov units of length and time, which means the non-dimensionalized Kolmogorov length $\eta$, time $\tau_{\eta}$, velocity $v_{\eta}$, and acceleration $a_{\eta}$ are unity) equation of motion reads,
\begin{equation}
	\begin{split}
		\ddot{\mathbf{x}} _p = \beta \frac{\mathrm{D}\mathbf{u}_f}{\mathrm{D}t} -\frac{1}{\text{St}}( \dot{\mathbf{x}}_p - \mathbf{u}_f) + 
		% (1- \frac{\beta}{3}) 
		a_{\text{e}} \sin(\omega t )\hat{\mathbf{e}}_x,
	\end{split}
	\label{eq1}
\end{equation}
where $\mathbf{u}_f$ is the flow velocity at the position of the particle, $\mathbf{v}_p$ is the instantaneous particle velocity $\mathbf{v}_p = \dot{\mathbf{x}}_p$, with $\mathbf{x}_p$ the particle position. Here $\beta = \frac{3}{1+2\rho_p / \rho_f}$ with $\rho_p / \rho_f$ the density ratio between the particle and the fluid, and $\text{St} =d_p^2/(12 \beta \nu \tau_{\eta})$ is the Stokes number of the particle of diameter $d_p$ and with kinematic fluid viscosity $\nu$. 

We perform three-dimensional fully resolved DNS of Navier-Stokes equations with light particles \cite{peyret2002spectral,bec2006acceleration,Calzavarinibook} (details provided in the Supplementary Information, {\color{blue}{\textit{SI Appendix}} Section A}). In the main paper, we focus on very light particles in the limit of $\beta \rightarrow 3.0$ ($\rho_p \rightarrow 0$, namely bubbles) and $\text{St}=1.0$ with resolution $N^3=256^3$. 
%, corresponding to $\text{Re}_{\lambda} \simeq 36$. 
Here $\text{St}=1.0$ is chosen such that light particles (with no external forcing applied) in HIT present strong preferential clustering in the vortex filaments \cite{maxey1994simulation,calzavarini2008dimensionality, mathai2020bubbly} and thus create a strong state of dynamic localization. 
The amplitude, $a_{\text{e}}$, and angular frequency, $\omega$, of the periodic external force are explored in a wide range with $a_{\text{e}} \in [1.0, 150.7]a_\eta$ and $\omega \in [0.2, 7.0]\tau_\eta^{-1}$. \textcolor{black}{These parameter ranges are chosen in such a way to explore all relevant physical time scales of the external forcing which may be bigger, shorter, or comparable to those of the vortex filaments (i.e., Kolmogorov time scales), so as} to provide a comprehensive understanding of light particle behaviors under external forcing.
We primarily focus on two aspects of the simulation results: 
(1) Temporal evolution features: This concerns instantaneous Lagrangian characteristics of individual particles. 
(2) Collective response of particle Lagrangian features: This refers to the ensemble-averaged or particle-averaged behaviors of particle motion. Here, particles are treated as a collective entity, and their macroscopic averaged responses are analyzed.  
Specifically, the ensemble average of a parameter $f$ of the $i^{th}$ particle at time $t$ is calculated as $\left<f \right>= \left< \left< f_i(t) \right>_{p}\right>_{t} = \frac{1}{n_p(T -t_0)}\int_{t_0}^{T}\sum_{i=1}^{n_p}f_i(t) dt$, where $\left< \cdot \right>$, $\left< \cdot \right>_p$, and $\left< \cdot \right>_t$ denote ensemble average, particle average, and time average, respectively. The time period $t_0$ to $T$ corresponds to the final statistical equilibrium state, and $n_p$ is the total number of particles. An equilibrium state is considered reached after two integral times, $T_L$ (for details see {\color{blue} \textit{SI Appendix} Fig.~S1}).
It is noted that statistically one particle dynamics (Lagrangian properties for a long time) and averaging over many particles (statistical ensemble) are equivalent \cite{allen2017computer}. 

Additional simulations are performed at different resolutions (see {\color{blue}\textit{SI Appendix} Section D}) to demonstrate the robustness of our findings, i.e., the revealed mechanism and the validity of the proposed theoretical model, across Reynolds numbers. Furthermore, we vary the characteristics of the particles used, spanning different combinations of their $\text{St}$ and $\beta$ (see {\color{blue}\textit{SI Appendix} Section E, and F}). We demonstrate that the resonant phenomenon is unique to light particles with $\beta$ values extending beyond $\beta \gtrsim 2$, since only light particles display trapping in vortex filaments.

\section*{Potential landscape experienced by light particles \\and their Lagrangian trajectories in HIT}

We start by examining the behavior of light particles in HIT without the influence of external forces. When light particles dive into turbulent flows, they become entrained within the core of vortex filaments. We will describe this trapping behavior as resulting from an effective ``potential landscape'' set by the vortex filament structure. Indeed, since $\mathrm{D}\mathbf{u}_f/\mathrm{D}t \sim \nabla P$ in Eq.~\ref{eq1}, we can heuristically understand that pressure plays the role of the effective potential landscape felt by the light particles.
At the core of a vortex, locally the pressure reaches a minimum, accompanied by a high enstrophy level (denoted by $\Omega = (\nabla \times \mathbf{u}_f)^2$, which is a measure for assessing the level of turbulence and the intensity of rotational motion in turbulence), see Fig.~\ref{FIG1}. 
The spatial distribution of light particles, depicted by filamentary-shaped clusters colored according to enstrophy at particle locations, corroborates their entrainment within vortex cores (particles colored red) without external force influence.

Next, we enrich the picture by introducing an external force as demonstrated in Fig.~\ref{FIG1}(e). A periodic external force oriented in the $x$ direction, $\mathbf{f}_E = a_{\text{e}} \sin(\omega t) \hat{\mathbf{e}}_x$, is applied directly on the light particles. 
One particle's instantaneous state can be either trapped or free from the vortex (sketched in Fig.~\ref{FIG1}(d)).
At varying $a_{\text{e}}$ (with $\omega$ fixed), three distinct regimes for the Lagrangian trajectory of light particles are identified: the localized (low $a_{\text{e}}$), the transitional (intermediate $a_{\text{e}}$), and the delocalized (high $a_{\text{e}}$) (see Fig.~\ref{FIG1}(f), with different colors representing the enstrophy level along the particle path). 
Different regimes are identified by different trajectory patterns and different particle oscillation extent along $x$ direction (in which the external force is applied). When $a_{\text{e}}$ is low, the particle trajectory exhibits a spiral motion pattern and drifting with vortices, with oscillations limited to a small range along the $x$ axis of the forcing, indicating a dominance of the vortex potential landscape.
As $a_{\text{e}}$ increases to intermediate levels (transitional regime), the particle trajectory combines a spiral motion and oscillation over a wider spatial range along the $x$ direction.
As $a_{\text{e}}$ reaches high levels (delocalized regime), 
the particle displays pronounced periodic oscillations extending over a large range along the $x$ direction. 
All these regimes reveal different levels of trap (warm colors in trajectory)-free (cold colors in trajectory) nature of particle motion in turbulence.
Further quantitative classification will be discussed in the next section.

\section*{Detailed Lagrangian characteristics}
\begin{figure*}[t!]
	\centering
	\includegraphics[width=.9\linewidth]{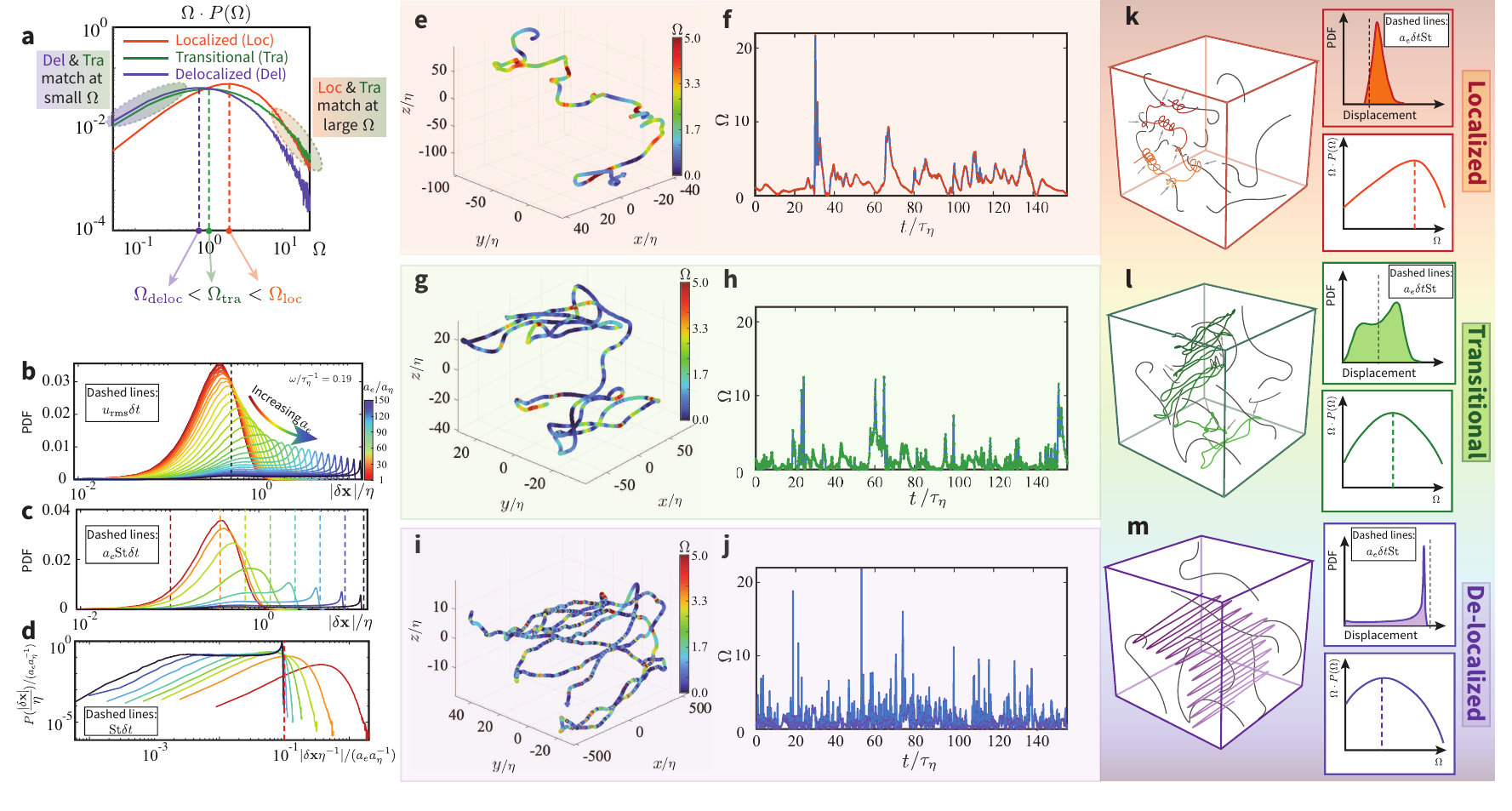}
	\caption{ {\bf Lagrangian characteristics of the localized, transitional, and delocalized regimes.} 
		{(a)} The normalized probability function of enstrophy, $\Omega P(\Omega)$, felt by the particles along their trajectories. {\textcolor{black}{The dashed vertical lines show the enstrophy corresponding to the peak of probability function. Note that the relative magnitudes of $\Omega_{\text{deloc}}$, $\Omega_{\text{tra}}$, and $\Omega_{\text{loc}}$ are merely a qualitative distinction among the three regimes.}}
		{(b)} PDF of the magnitude of displacement of the particles, under unit time increment ($\delta t \sim 0.1 \tau_\eta$). With increasing $a_{\text{e}}$ (denoting by the colormap, and here one typical $\omega$, i.e., $\omega/\tau^{-1} = 0.19$ is chosen), the PDF of $|\delta \mathbf{x}|$ displays (1) obviously single broad peak for low $a_{\text{e}}$, with the peak located at $|\delta \mathbf{x}| \sim u_{\textrm{rms}}\delta t$ which is induced by the large scale velocity sweeping the vortex filaments around, (2) double peak for intermediate $a_{\text{e}}$, and (3) single low peak for high $a_{\text{e}}$ with the peak located at $|\delta \mathbf{x}| \sim a_{\text{e}} \text{St} \delta t$, where $a_{\text{e}} \text{St}$ is the free-fall velocity that is determined by the balance between the drag force and the external force. These three identified features of the PDFs correspond to the localized, transitional, and delocalized regimes, respectively. 
		{(c)} The free-falling displacement, $a_{\text{e}}\text{St}\delta t$ sets an upper bound on $\delta \mathbf{x}$ for the delocalized case (purple solid line) where $\mathbf{f}_E$ plays a major role. While for the localized and for the transitional cases, $\mathbf{f}_p$ is the leading factor or is balanced by $\mathbf{f}_E$. 
		As the curves in {(c)} are normalized by $a_{\text{e}}$, all the dashed lines collapse (see {(d)}). 
		{(e)}The zoom-in view on the particle trajectory in the localized regime. 
		{(f)} The instantaneous enstrophy corresponding to the process of the localized particle travelling in {(e)}.
		% and the PDF of $|\delta \mathbf{x}|$ is attached in the inset. 
		Figures similar to {(e)} - {(f)} are also reported for the transitional regime ({(g)} - {(h)}) and for the delocalized regime ({(i)} - {(j)}). 
		(k-m) Summary of the main characteristics of the three regimes. The sketches demonstrate the main features of the trajectory, PDF of displacement magnitude $|\delta \mathbf{x}|$, and the normalized PDF of enstrophy, $\Omega P(\Omega)$ of each regime. 
	}
	\label{FIG2}
\end{figure*}

\begin{figure*}%[tbhp]
	\centering
	\includegraphics[width=.99\linewidth]{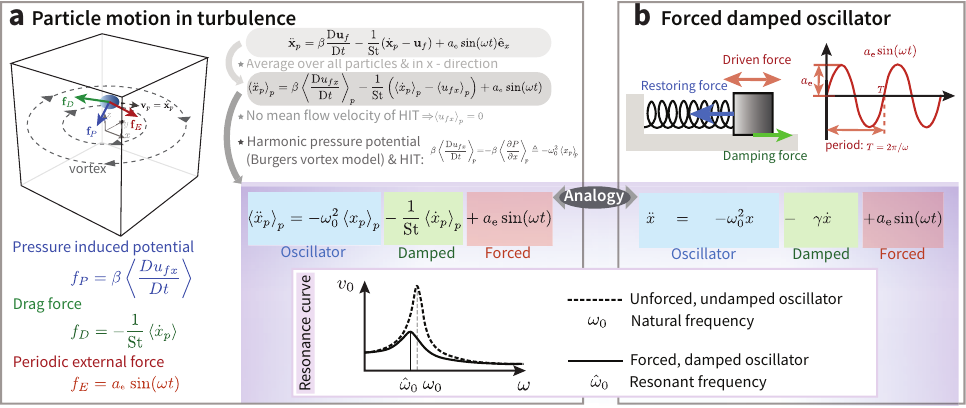}
	\caption{{\bf Analogue model of particle motion in turbulence.} {(a)} Particle motion in turbulence, considering one-way coupled point particles experiencing drag force ($\mathbf{f}_D$), the force due to the net effect of added mass and pressure ($\mathbf{f}_P$), and a periodic external force $\mathbf{f}_E$. Lift, buoyancy, two-way coupling, and particle-particle interactions are neglected. The equation of particle motion is nondimensionalized by Kolmogorov units of length ($\eta$) and time ($\tau_{\eta}$). Corresponding particle-averaged equations in the x direction are derived.
		{(b)} Forced damped harmonic oscillator, depicting a spring oscillator influenced by restoring (elastic) force, damping, and periodic external driven force (same as (a)). The blue, green, and red shaded boxes show the resemblance between the particle motion in turbulence and the forced damped harmonic oscillator with respect to the oscillator, damped, and forced terms. 
		As the external angular frequency, $\omega$, equals the natural frequency (resonant frequency), $\omega_0$ ($\hat{\omega}_0$), corresponding to an unforced, undamped oscillator (forced damped oscillator), the system reaches resonance, leading to maximum velocity oscillation amplitude (depicted by the resonance curve).
	}
	\label{FIG3}
\end{figure*}

By closely observing the enstrophy (represented by colors in Fig.~\ref{FIG1}(f)) for each trajectory of a particle in each regime, it becomes clear that the particle in different regimes travels through different regions of the turbulent field. To quantify this, we plot the compensated probability density function of the enstrophy, $\Omega P(\Omega)$, felt by the particles along their trajectories (Fig.~\ref{FIG2}(a)).
This shows a relative preference of particles for different regimes; specifically, in the localized regime, particles tend to spend more time in the vortex (high enstrophy level region), while those in the delocalized regime prefer to roam freely outside of the vortex (low enstrophy level region). 
It is noteworthy that the PDF of enstrophy in the transitional regime combines the features of the localized and delocalized regimes: it follows delocalized (localized) regime for weak vortex structures (small $\Omega$), while it still remains trapped by the strongest vortex structures, following the Localized regime at large $\Omega$ (highlighted by the shaded areas in Fig.~\ref{FIG2}(a)).

To better assess whether the behavior of particles is predominantly influenced by vortices or by the external force, we focus on the statistics of the particle displacement (magnitude) per unit time ($\delta t \sim 0.1 \tau_\eta$), and $|\delta \mathbf{x}| = \left<|\mathbf{x}_p(t+\delta t) - \mathbf{x}_p(t)| \right>$. Fig.~\ref{FIG2}(b) displays the PDF of $|\delta \mathbf{x}|$ when varying $a_{\text{e}}$ (with a fixed typical value of $\omega$). 
As $a_{\text{e}}$ increases, the PDF of $|\delta \mathbf{x}|$ exhibits distinct features: (1) a single broad peak for low $a_{\text{e}}$, positioned at $|\delta \mathbf{x}| \sim u_{\text{rms}}\delta t$, induced by the large scale velocity sweeping the vortex filaments around, where $u_{\text{rms}}=(\sum_{i = x,y,z}u_i^2)^{1/2}$ characterizing turbulent velocity fluctuations (shown by dashed lines in Fig.~\ref{FIG2}(b)); (2) double peaks for intermediate $a_{\text{e}}$; and (3) a single sharp peak for high $a_{\text{e}}$, located at $|\delta \mathbf{x}| \sim |\delta \mathbf{x}_{\text{free}}|$, where $|\delta \mathbf{x}_{\text{free}}|=a_{\text{e}} \text{St} \delta t$ represents the free fall velocity determined by the balance between drag force ($\mathbf{f}_{D}$), and external force ($\mathbf{f}_{E}$). \textcolor{black}{These features, corresponding to the localized, transitional, and delocalized regimes, respectively, can be used to classify qualitatively the three regimes}. The displacement by the free-fall particle, $|\delta \mathbf{x}_{\text{free}}|$, sets an upper bound for $\delta \mathbf{x}$ in the delocalized regime (purple solid line), where $\mathbf{f}_E$ predominates. While for the localized and transitional regimes, $\mathbf{f}_P$ is the leading factor or is comparable to $\mathbf{f}_E$, such that particles can exceed their freefall velocity due to turbulent fluctuations.

Fig.~\ref{FIG2}(e-m) shows typical trajectories in the three different regimes. We observe the trap-free dynamics set by the vortex filaments. While in the localized regime, the particles spend a significant amount of time in the high enstrophy regions, travelling from one vortex to the next, this reduces in the transitional regime, until finally, in the delocalized regime, the particles are just skimming through the vortex landscape, showing very short bursts of high enstrophy as it happens to pass through a filament.

\section*{Particle motion in turbulence: analogue forced damped oscillator}
Understanding the nature of the different forces that govern the dynamics of the forced light particles in the different regimes, we can make a useful analogy to one of the most classical systems in physics: the forced damped harmonic oscillator.

In the previous section, we have shown the full equation of motion of a light particle subjected to a periodic force in HIT (see Eq.~\ref{eq1}). We can then perform the particle average of Eq.~\ref{eq1} and only consider the particle motion in the $x$ direction in which the external force is applied. This gives
\begin{equation}
	\begin{split}
		\left<\ddot{x}_p\right>_p = \beta \left<\frac{\mathrm{D} {u}_{fx}}{\mathrm{D}t}\right>_p  -\frac{1}{\text{St}} \left(\left<\dot{ {x}}_p\right>_p -\left<u_{fx}\right>_p \right) + a_{\text{e}} \sin(\omega t ).
	\end{split}
	\label{eq2}
\end{equation}

\textcolor{black}{As the current investigated flow is turbulent (high Reynolds number flows), the Navier-Stokes equation approaches the Euler equation where the material derivative of the fluid velocity is dominated by the pressure gradient, and the viscous and forcing terms can be neglected, and thus the first term on the right hand side has a good approximation of $\left<\mathrm{D} {u}_{fx}/ \mathrm{D}t \right>_p \sim \left<\partial P/\partial x \right>_p$ \cite{clift2013bubbles}. This pressure-induced force acts as the restoring force of the particle oscillation within the vortex filaments and can be connected to the natural resonance frequency (as discussed later). The friction experienced by the particle as it moves in the fluid is explicitly captured by the drag force on the particle, which is considered in the ``damping term'' (the second term on the right-hand side).}

Previous works \cite{jimenez1993structure, jimenez1998characteristics, kang2007dynamics, ghira2022characteristics} have reported that the small-scale vortex structures can be represented by a classical Burgers vortex model \cite{burgers1948mathematical}. This model implies a harmonic pressure potential with respect to the position, and we assume the vortex filaments are axisymmetric, so the pressure can locally be expressed as $P = P_0 + \alpha|\mathbf{x}_p-\mathbf{x}_{vf}|^2
$ with $\alpha$ being a scaling factor and $|\mathbf{x}_p-\mathbf{x}_{vf}|$ the magnitude of the relative displacement of a particle with respect to the vortex core at $\mathbf{x}_{vf}$. 
The effective potential well set by this pressure field induces a force that is, in the considered $x$ direction, given by $\frac{\partial P}{\partial x} = 2\alpha(x_p - x_{vf})$, and because the turbulence is homogeneous and isotropic we have $\left<x_{vf}\right>_p = 0$. So we can simplify the term of the net effect of added mass and pressure as  
\begin{equation}
	\begin{split}
		\beta \left< \frac{\mathrm{D}u_{fx}}{\mathrm{D}t} \right>_p \sim -\beta \left<\frac{\partial P}{\partial x}\right>_p
		% &= -2\alpha \beta \left<x_{p} - x_{vf}\right>_p\\
		% &= -2\alpha \beta \left<x_{p}\right>_p\\
		\triangleq -\omega_0^2 \left<x_{p}\right>_p,
	\end{split}
	\label{burgers}
\end{equation}  
with $\omega_0 = 2\alpha \beta$ being the average natural frequency of the vortex filaments. \textcolor{black}{Note that $x_p$ needs to be accounted for in unwrapped coordinates.}

Next we consider the term $\left< u_{fx} \right>_p$, which is the Eulerian flow velocity at the position of the particle in Eq.~\ref{eq2}. Crucially, due to the particle averaging, we obtain $\left< u_{fx} \right>_p = 0$ due to the isotropy of the background turbulence. The particle averaging thus acts as a mechanism to disentangle the collective motion in response to the external force from the local turbulent fluctuations (see {\color{blue}{\textit{SI Appendix}} Section B, Fig.~S3(a)-(b)}).
Based on the aforementioned arguments (detailed discussion is shown in {\color{blue}{\textit{SI Appendix}} Section B}) and emulating the form of the equation for the forced damped oscillator, we can reorganize Eq.~\ref{eq2} as,
\begin{equation}
	\begin{split}
		\left<\ddot{x}_p\right>_p = -\omega_0^2 \left<{x}_p\right>_p -\gamma \left<\dot{ {x}}_p\right>_p  + a_{\text{e}} \sin(\omega t ),
	\end{split}
	\label{final}
\end{equation}
with $\gamma = 1/{\text{St}}$ relating the damping force in turbulence. 
Since the time derivative and particle averaging operators commute, we have thus mapped the dynamics of the collective particle motion $\langle x_p \rangle$ onto that of a forced damped harmonic oscillator with natural frequency $\omega_0$, damping $\gamma$, forcing frequency $\omega$ and forcing amplitude $a_e$. This comparison is illustrated in Fig.~\ref{FIG3}(a)-(b).
This analogy suggests the possibility to find resonant behavior in the collective particle dynamics. For an undamped oscillator, resonance occurs when the external force angular frequency, $\omega$, matches the natural frequency, $\omega_0$. However, in a forced damped oscillator, the resonant frequency, $\hat{\omega}_0$, is slightly modulated by the damping (see resonance curve in Fig.~\ref{FIG3}).
It is important to note that, in the scenario of particle motion in turbulence, $\omega_0$ reflects the intrinsic properties of the background turbulence: the ``curvature'' of the pressure field, i.e. the depth of the effective potential well. 

\begin{figure*}[t!]
	\centering
	\includegraphics[width=0.8\linewidth]{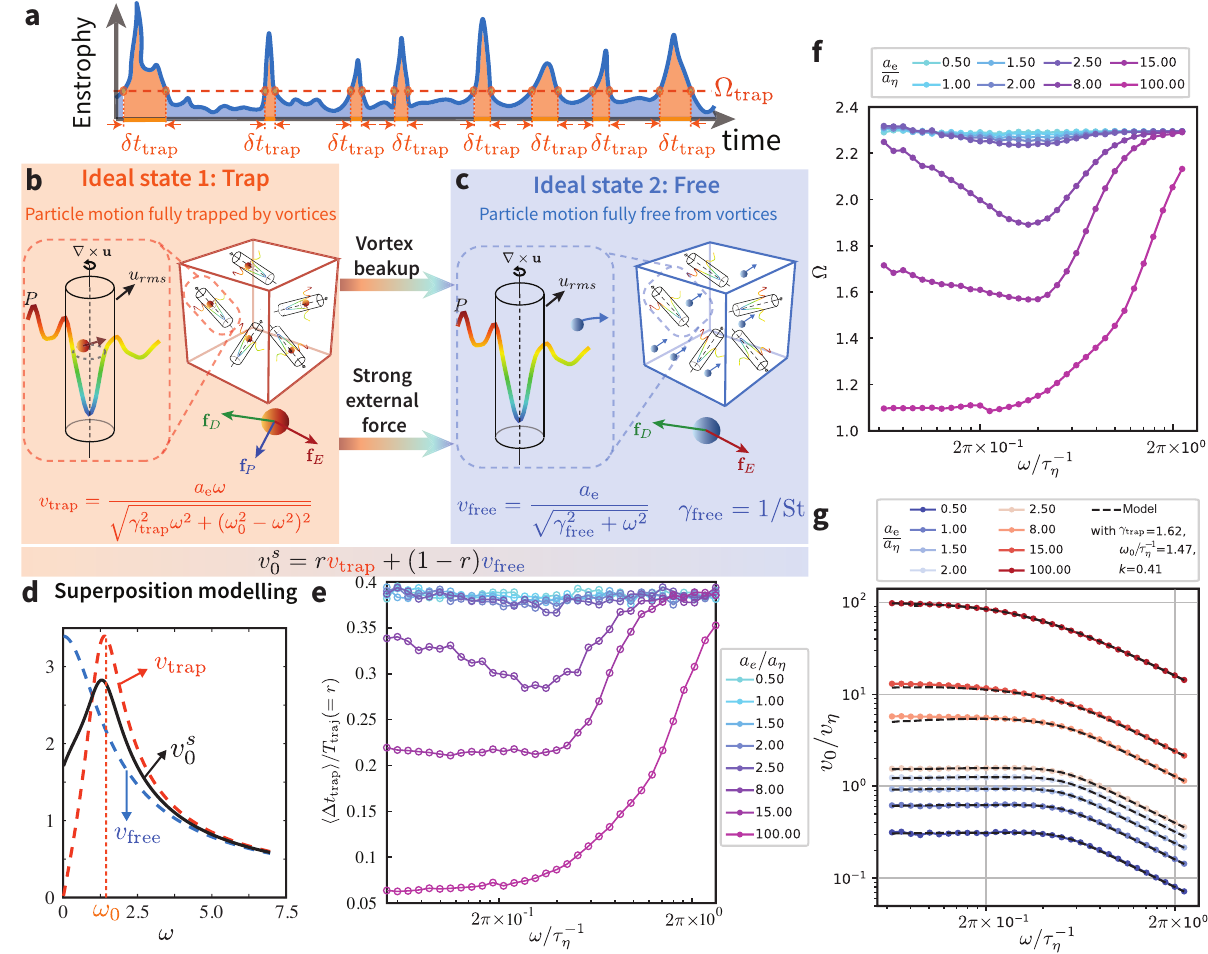}
	\caption{{\bf Superposition modelling.}
		{(a)} Definition of the localized interval, $\delta t_{\text{trap}}$ (represented by the orange shaded area), along the particle travelling process. $\Omega_{\text{trap}}$ is the threshold above which the particle is regarded as trapped by the vortex (orange shaded range).
		(b) and (c) describe two ideal states: ``trap'' and ``free''. 
		The ``trap'' state is when all the particles are trapped within dynamic potential wells created by the vortices and their collective mobility is low. The particle behavior resembles a forced damped oscillator (under the influence of $f_D$, $f_P$, and $f_E$), described by control equation Eq.~\ref{final}
		The velocity oscillation amplitude is $v_{\text{trap}}$.
		The ``free'' state is when the particle motion is delocalized from the vortices and freely travelling in the turbulent flow (under the influence of ${f}_D$ and ${f}_E$), governed by control equation Eq.~\ref{forced_damped_eqn}. The particle motion is only minimally influenced by the vortices and their collective mobility is high. 
		The amplitude of the velocity oscillation is $v_{\text{free}}$.
		The real particle motion (Fig.~\ref{FIG4}(a)) is a combination of the ``trap'' state (orange shaded range) and the ``free'' state (blue shaded range). 
		Therefore, the collective amplitude of the velocity oscillation, $v_0$, can be modelled as the superposition of the two ideal states: $v_0^s = rv_{\text{trap}}+(1-r)v_{\text{free}}$. 
		The superposition ratio $r$ bears the physical meaning of the fraction of localized time, which can be calculated by $r=\left<\Delta t_{\text{trap}}\right>_p/T_{\text{traj}}$, where $\Delta t_{\text{trap}} = \sum{\delta t_{\text{trap}}}$ is the proportion of the total localized time interval and $T_{\text{traj}}$ represents the total particle trajectory duration.
		(d) The superposition velocity amplitude, $v_0^s$ (black line), is deduced by applying a prefactor $r$ as the superposition ratio between the two ideal velocity amplitude solutions $v_{\text{trap}}$ (orange dashed line) and $v_{\text{free}}$ (blue dashed line).
		{(e)} The proportion of localized time, $\left<\Delta t_{\text{trap}}\right>_p = \sum{\delta t_{\text{trap}}}$, within the total particle travelling duration, $T_{\text{traj}}$. 
		(f) Enstrophy level felt by the light particles at their positions as a function of $a_{\text{e}}$ and $\omega$. Here $\Omega = \left< \Omega(\mathbf{x}_p, t)\right>$.
		(g) \textcolor{black}{Comparison of the particle averaged amplitude of oscillating velocity (normalized by Kolmogorov velocity scale $v_{\eta}$), $v_0/v_{\eta}$, between the superposition modelling results (dashed lines) and the simulation results (symbols)}
		(details of calculating $v_0$ can be found in {\color{blue}\textit{SI Appendix} Section C}).
		The prefactor $r$ is a mapping from the enstrophy by a linear transformation, i.e., $r (a_{\text{e}}, \omega) = k \tilde{\Omega}(a_{\text{e}}, \omega)$ and $k$ is a constant.
		The parameters in the superposition model are $\gamma_{\text{trap}} = 1.62$, $\omega_0 = 1.47$, and $k=0.41$. $\tilde{\Omega}$ is normalized by $\tilde{\Omega}(a_{\text{e}}, \omega) = (\Omega(a_{\text{e}}, \omega) -\Omega_{\text{free}})/(\Omega_{\text{trap}} - \Omega_{\text{free}})$, where $\Omega_{\text{trap}}$ and $\Omega_{\text{free}}$ correspond to the high and low enstrophy plateaus in (f) respectively, depending on $Re_{L}$ only. 
	}
	\label{FIG4}
\end{figure*}

Before we can seek this resonance in the data, we need to refine our model of the collective dynamics a bit more. The solution to Eq.~\ref{final} is (hereafter we drop the particle averaged symbol $\left<\cdot\right>_p$ for simplicity) 
\begin{equation}
	x_p(t) = x_0 \sin(\omega t),
	\label{forced_damped_oscillator_sol}
\end{equation}
where $x_0$ is the oscillation amplitude of the position, $x_0 = a_{\text{e}} / \sqrt{\gamma^2\omega^2 + (\omega_0^2-\omega^2)^2}$.

Computing the time derivative of Eq.~\ref{forced_damped_oscillator_sol}, we can get the velocity, 
\begin{equation}
	v(t) =  \frac{a_{\text{e}}\omega}{\sqrt{\gamma^2\omega^2+(\omega_0^2 - \omega^2)^2}} \cos(\omega t).
	\label{forced_damped_oscillator_v}
\end{equation}

Note that Eq.~\ref{final} is applicable only when the particle is in a state completely trapped by the vortex, i.e., during those time intervals the enstrophy is above the threshold $\Omega_{\text{trap}}$ (see Fig.~\ref{FIG4}(a)). However, there are still instances when the particle experiences free travel, strongly driven by the external force and thus escaping from the vortex or being freely released by vortices due to their finite lifetimes. Under this condition, the equation of motion simplifies to:
\begin{equation}
	\begin{split}
		\ddot{x}_p= -\gamma \dot{ {x}}_p  + a_{\text{e}} \sin(\omega t ).
	\end{split}
	\label{forced_damped_eqn}
\end{equation}
Correspondingly, the velocity is
\begin{equation}
	v(t) =  \frac{a_{\text{e}}}{\sqrt{\gamma^2+\omega^2}} \cos(\omega t).
	\label{forced_damped_v}
\end{equation}

To fully model the collective particle dynamics, we thus need to combine both the forced damped harmonic oscillator solution Eq.~\ref{forced_damped_oscillator_v}, as well as the free solution Eq.~\ref{forced_damped_v}. We will refer to this as the superposition model, where the total velocity amplitude $v_0^s$ is defined as a combination of two ideal states
\begin{equation}\label{eq:super}
	v_0^s = r v_{\text{trap}} + (1-r) v_{\text{free}}.
\end{equation}

Here $v_{\text{trap}} = a_{\text{e}}\omega/\sqrt{\gamma_{\text{trap}}^2\omega^2+(\omega_0^2 - \omega^2)^2}$ is the velocity amplitude in Eq.~\ref{forced_damped_oscillator_v} when the particle is trapped in the vortex filament. On the other hand $v_{\text{free}} = a_{\text{e}}/\sqrt{\gamma_{\text{free}}^2+\omega^2}$ is the velocity amplitude in Eq.~\ref{forced_damped_v} when the particles are freely travelling outside the vortex filaments with minimal collective influence of the background turbulence, referred to as the ``free'' state. \textcolor{black}{The damping effect induced by the fluid is different in these two states, indicated by different damping ratios $\gamma_{\text{trap}}$ and $\gamma_{\text{free}}$. Based on the superposition model, we find that $\gamma_{\text{free}} \approx 1/St$, and $\gamma_{\text{trap}}$ is slightly larger than $1/St$ (as the Reynolds number varies $\gamma_{\text{trap}}$ remains relatively stable with $\gamma_{\text{trap}} \approx 1.65 \pm 0.08$, indicating the robustness of the superposition model). This is due to the different fluid flow conditions in or outside the vortex filaments. When the particle is trapped inside the vortex, the shearing fluid induces a higher friction effect (corresponding to a larger $\gamma_{\text{trap}}$). It can be understood as an enhanced eddy friction in the high vorticity region. While the particle is outside the vortex and/or when it is highly forced, it approaches the friction of a particle moving in an otherwise quiescent domain, yielding $\gamma_{\text{free}} \approx 1/St$, which agrees with what we get from the interpretation of the simulation results with the theoretical model.} The ratio $r$ physically represents the fraction of time (or equivalently, the fraction of particles \textcolor{black}{at any instant}) spent in the ``trap'' state, see Fig.~\ref{FIG4}. \textcolor{black}{The value of $r$ varies with the strength of the trapping potential posed by the vortex in turbulence, influenced by factors such as vortex structure features (e.g., hairpin-like structures in wall-bounded turbulent flows) and turbulent intensity (i.e., Reynolds number, as shown in {\color{blue}\textit{SI Appendix} Fig. S4 (d)} and \textcolor{blue}{Fig. S5 (d)}.}

The fitting procedure of this model is laid out in {\color{blue}{\textit{SI Appendix}} Section B}. \textcolor{black}{In brief, the superposition modelling optimizes three free parameters, describing the damping, the natural frequency and the proportion of trapped particles, using the simulation results. This model predicts the average particle behavior across a wide range of conditions, with parameters remaining stable as the Reynolds number varies, demonstrating the robustness of the model.} The results are presented in Fig.~\ref{FIG4}(g), showing an excellent agreement between the collective velocity amplitude obtained from the DNS (symbols) and our superposition model Eq.~\ref{eq:super} (dashed lines).

\begin{figure*}[t!]
	\centering
	\includegraphics[width=.97\linewidth]{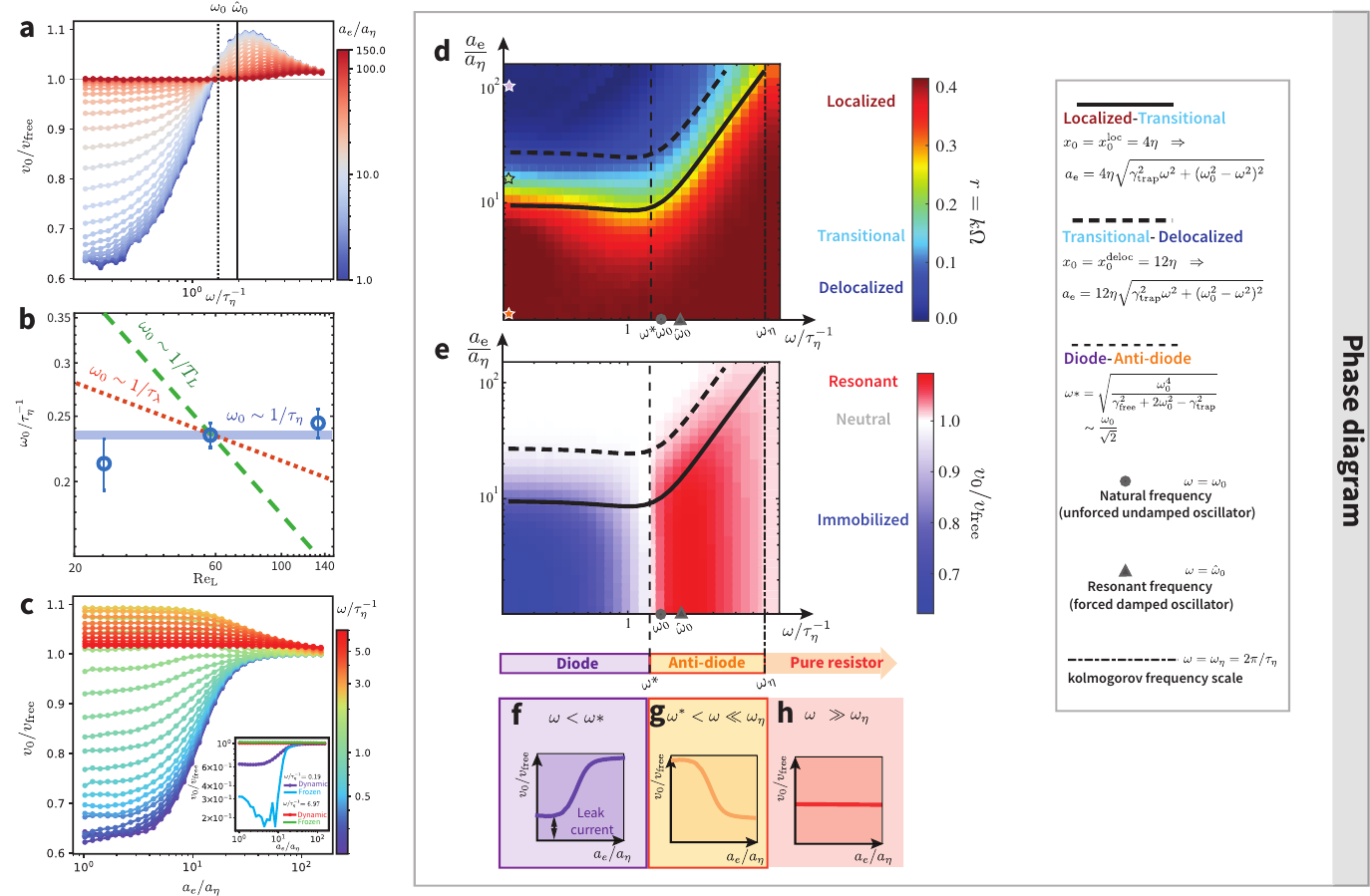}
	\caption{{\bf phase diagram.} 
		{(a)} Resonance curve $v_0/v_{\text{free}}$ as a function of $\omega$ and $a_{\text{e}}$. 
		% The thick lines connecting the symbols are drawn to guide the eye. 
		The dotted and solid vertical lines are $\omega_0$ and $\hat{\omega}_{0}$ respectively, corresponding to natural frequency (unforced undamped oscillator) and resonant frequency (forced damped oscillator).
		(b) Reynolds number dependence of the natural frequency, $\omega_0$. Symbols are based on varying Re$_{\text{L}}$, whose errorbar is estimated by applying different $a_{\text{e}}$ while ensuring $r>0.35$. Green dashed line and red dotted line show scaling characteristics of $\omega_0 \sim 1/T_{L} \propto \text{Re}_{\text{L}}^{-1/2}$ and $\omega_0 \sim 1/\tau_{\lambda} \propto \text{Re}_{\text{L}}^{-1/6}$, respectively. The blue thick line shows $\omega_0 \sim 1/\tau_{\eta} \propto \text{Re}_{\text{L}}^{0}$.
		{(c)} Particle velocity oscillation amplitude as a function of $a_{\text{e}}$ and $\omega$. The particle velocity oscillation amplitude, $v_0$, is normalized by the theoretical free solution, $v_{\text{free}}$. 
		Unlike the diode, the particles still display some mobility in the smaller $a_{\text{e}}$ range, indicating some ``leaky current'' under weak driving force, which is induced by the finite lifetime of the vortex. 
		This can be verified by the simulation of particle motion in a frozen Eulerian flow field, where we take a snapshot of the Eulerian flow field at the statistically stationary state and enforce it to remaine unchanged. The leaky current drops a lot (but not equal to zero due to the finite-sized effect of vortices, shown by the blue line in the inset). 
		The inset shows the comparison between the dynamic field and the frozen field of two typical cases of $\omega= 0.19$ and $\omega = 6.97$.
		(d) A phase diagram for the dynamics of a light particle influenced by a periodic external force in HIT. 
		The phase diagram is colored by different levels of the superposition ratio (fraction of localized time) $r$ ($=k\Omega$). 
		Three identified regimes are localized, transitional, and delocalized regimes.
		The solid and dashed thick lines indicate the boundary between different regimes.
		$x_0^\text{loc}$ ($\sim 4 \eta$) is the upper bound of the oscillating amplitude of the particle position in the localized regime. $x_0^\text{deloc}$ ($\sim 12 \eta$) is the lower bound of the oscillating amplitude of the particle position in the delocalized regime. A transitional regime is identified when the particle oscillates in the range from $x_0^\text{loc}$ to $x_0^\text{deloc}$. 
		The values of $x_0^\text{loc}$ and  $x_0^\text{deloc}$ are determined by fitting the simulation results. The stars
		(from low $a_{\text{e}}$ to large $a_{\text{e}}$) marked in the diagram indicate three typical cases from each regime corresponding to those in Fig.~\ref{FIG2}(e), (g), and (i). 
		The dashed vertical line is $\omega=\omega^*$ whose physical meaning will be explained in (e).
		The gray circle on $x$-axis marks $\omega = \omega_0$.
		The gray triangle on $x$-axis marks $\omega=\hat{\omega}_0$.
		The dash-dotted vertical line is $\omega=\omega_{\eta}$ and $\omega_{\eta}=2 \pi/\tau_\eta$.
		(e) A phase diagram for the response mobility, $v_0/v_{\text{free}}$, of a light particle influenced by a periodic external force in HIT.  
		The regime space is colored by different levels of $v_0/v_{\text{free}}$. 
		Three identified regimes are the immobilized, resonant, and neutral.  
		To highlight the main feature of the characteristic curve under different $\omega$ ranges (see (e)), three regimes are revealed in (f)-(h). 
		(f) Diode mode: $\omega < \omega^*$, 
		(g) ``Anti-diode'' mode: $\omega^*<\omega \ll \omega_{\eta} $). 
		(h) Pure resistor mode: $\omega \gg \omega_{\eta}$. 
		The diode and anti-diode boundary is marked by the thin vertical dashed line, which corresponds to $\omega* = (\omega_0^4/{(\gamma_{\text{free}}^2 + 2\omega_0^2-\gamma_{\text{trap}}^2)})^{1/2}$, and when $\gamma_{\text{trap}} \approx \gamma_{\text{free}}$ we have $\omega*\sim {\omega_0/\sqrt{2}}$. Detailed calculation of the boundaries between different regimes and modes are reported in {\color{blue}\textit{SI Appendix} Section A}
	}
	\label{FIG5}
\end{figure*} 

%To fully model the collective particle dynamics, we thus need to combine both the forced damped harmonic oscillator solution Eq.~\ref{forced_damped_oscillator_v}, as well as the free solution Eq.~\ref{forced_damped_v}. We will refer to this as the superposition model, where the total velocity amplitude $v_0^s$ is defined as a combination of two ideal states
%\begin{equation}\label{eq:super}
%	v_0^s = r v_{\text{trap}} + (1-r) v_{\text{free}}.
%\end{equation}
%Here $v_{\text{trap}} = a_{\text{e}}\omega/\sqrt{\gamma_{\text{trap}}^2\omega^2+(\omega_0^2 - \omega^2)^2}$ is the velocity amplitude in Eq.~\ref{forced_damped_oscillator_v} when the particle is trapped in the vortex filament. On the other hand $v_{\text{free}} = a_{\text{e}}/\sqrt{\gamma_{\text{free}}^2+\omega^2}$ is the velocity amplitude in Eq.~\ref{forced_damped_v} when the particles are freely travelling outside the vortex filaments with minimal collective influence of the background turbulence, referred to as the ``free'' state. The ratio $r$ physically represents the fraction of time (or equivalently, the fraction of particles) spent in the ``trap'' state, see Fig.~\ref{FIG4}.
%
%The fitting procedure of this model is laid out in {\color{blue}{\textit{SI Appendix}} Section B}. The results are presented in Fig.~\ref{FIG4}(g), showing an excellent agreement between the collective velocity amplitude obtained from the DNS (symbols) and our superposition model Eq.~\ref{eq:super} (dashed lines).

\section*{Mechanism of localization-delocalization transition of light particles in turbulence.}
Finally, with the theoretical model in place, we can take a closer look at the DNS data and study the collective motion amplitude $v_0$ of the particle velocity in response to the varying external force, allowing us to classify the full phase space of the light particle motion, see Fig.~\ref{FIG5}.

As shown in Fig.~\ref{FIG5}(a), at low forcing amplitudes and low forcing frequency, we observed a decreased mobility of particles with respect to the free travelling mode due to the trapping of particles in vortex filaments.

Furthermore, as hypothesized from the forced damped harmonic oscillator analogy, we observe that as the forcing frequency is varied, there is a remarkable resonance between the particles and the vortex filaments observed in the collective motion of particles (represented by the normalized velocity oscillation amplitude $v_0/v_{\text{free}}$), around the resonant frequency denoted as $\hat{\omega}_0$ (with the vortex natural frequency denoted as $\omega_0$ see Fig.~\ref{FIG5}(a)). Around this resonance frequency, the particle velocity amplitude exceeds that of the free travelling mode. This resonance is robust for small and moderate forcing amplitudes, as exemplified by the collapse of the curves, but vanishes at large amplitude as we enter the delocalized regime and the influence of the background turbulence vanishes, alleviating the constraint by vortex filaments.

We find that the obtained natural resonance frequency $\omega_0$ scales most closely as the Kolmogorov frequency $\omega_{\eta}$ (see Fig.~\ref{FIG5}(b)). Physically, this is in agreement with the notion of resonance with vortex filaments with the typical turnover frequency of Kolmogorov scale.

We can also regard the collective particle velocity amplitude response as a function of the forcing amplitude as done in Fig.~\ref{FIG5}(c).
As the forcing amplitude is increased, the system undergoes the transition from the localized state at small forcing amplitude, where there is significant trapping by vortex filaments, to the delocalized state at large forcing amplitudes that yields the free travelling mode.
However, as we have seen, in the localized state, the particles can either display a decreased mobility due to their trapping or an increased mobility due to resonance, depending on the forcing frequency. As a consequence, different nonlinear responses can be identified upon gradually increasing $a_{\text{e}}$ (for a fixed $\omega$), whose changing rate (slope) of the response indicates the overall \textcolor{black}{mobility} properties of the system. 

For the different $\omega$, this response $v_0/v_{\text{free}}$\textcolor{black}{, induced by the driving external forcing,} displays different nonlinear trends as shown in Fig.~\ref{FIG5}(c), which are, by phenomenological classification similar to \textcolor{black}{the characteristic curves of current (response) with respect to the voltage (driving) of the} nonlinear electric elements, ``diode'' (\textcolor{black}{at a given small $\omega$, the system behaves like a diode when the response parameter $v_0/v_{\text{free}}$ increases as the driving parameter $a_e/a_{\eta}$ increases}, like a response switch-on), ``anti-diode'' (\textcolor{black}{at a given intermediate $\omega$, the system behaves like an anti-diode when the response parameter $v_0/v_{\text{free}}$ decreases as the driving parameter $a_e/a_{\eta}$ increases}, like a response shut-down), and ``pure resistor'' (response unchanged for large $\omega$), whose main features are highlighted in Fig.~\ref{FIG5}(f-h), respectively. Note, that in the immobilized state at small forcing frequency and amplitude, particles are not fully immobilized due to the finite lifetime of vortex filaments, which effectively yields a finite ``leaky current'' of escaping particles\textcolor{black}{, i.e., the response (particle velocity oscillation amplitude) is in fact non-zero while the driving factor (external forcing amplitude) reaches zero}.

The full phase diagram can then be summarized as a function of both forcing amplitude and frequency as presented in Fig~\ref{FIG5}(d-e). In Fig~\ref{FIG5}(d), we show the fraction $r$ of particles in the trapped state as obtained from the fit to the superposition model. This allows us to classify \textcolor{black}{quantitatively} the localized, transitional and delocalized state. As shown in Fig~\ref{FIG5}(e), within the localized regime, an immobilized mode ($v_0/v_{\text{free}}<1$) can be identified at low forcing frequencies $\omega \ll \omega_0$ due to the particle trapping, while at forcing frequencies close to Kolmogorov scales $\omega \approx \omega_0$, enhanced mobility is found due to the resonance with vortex filaments ($v_0/v_{\text{free}}>1$).

It is noteworthy to point out that we can employ the superposition model to obtain precise criteria for the boundaries between the regimes (localized, transitional, and delocalized) and different modes (immobilized to resonant to neutral). The transitions between the localized, transitional and delocalized regime are provided by isolines of the positional oscillation amplitude of the forced harmonic oscillator model (detailed calculation are reported in {\color{blue}\textit{SI Appendix} Section B}). We find that the transitional regime starts when the position oscillation amplitude exceeds $4\eta$, while the delocalized regime starts at a position oscillation amplitude of $12\eta$. Physically, we can understand this as being a proxy for the effective width of the potential well sensed by the light particles due to the vortex filaments, while the resonance frequency dictates the depth of the effective potential well. The transition between the immobilized and resonant modes is trivially obtained from the resonance frequency as fitted by the superposition model.

This reveals that one can employ the current understanding of the collective behavior of forced light particles in turbulence to probe the properties of the underlying turbulent flow itself. By searching for the resonant frequency of the bubble motion, one can identify the natural frequency $\omega_0$ of the vortex filaments in the underlying fluid, which relates to the Kolmogorov frequency of the turbulent flow. Furthermore, by studying the transition from the localized to the delocalized state, one can proxy the width of the vortex filaments in the underlying fluid. Indeed, the width that we find here for the vortex filaments in HIT, approximately $4\eta$, is comparable to those obtained through the identification and statistical analysis of vortex structures in a variety of turbulent ﬂows and for a wide range of Reynolds numbers in Eulerian-based statistical studies \cite{tanahashi2001appearance,kang2007dynamics,ghira2022characteristics,da2011intense,ganapathisubramani2008investigation}). 

\section*{Conclusions and Outlook}
We have provided a new tool based on the tight physical duality between the dynamics of light particles and the small-scale vortex in turbulence. By applying a modulated forcing on bubbles we have shown how bubbles can be employed to accurately investigate the statistical properties of small-scale structures in turbulence. Bubbles can be turned into accurate probes of a ``turbulence microscope''. On the other hand, the external force, leading to resonant effects between bubbles and vortex filaments, can trigger the delocalization of bubbles with a major impact on the average and collective macroscopic dynamics of bubbles in turbulence. 

At changing the properties of the external force, bubbles go through three distinct regimes identified as: localized, transitional, and delocalized.
In the localized regime, increasing $\omega$ (with $a_{\text{e}}$ unchanged), a remarkable resonant phenomenon emerges between particles and vortices, representing a mode transition from immobilized to resonant to neutral states, by which the natural frequency of the vortex, $\omega_0$, can be detected, i.e. $\omega_0 \sim \mathcal{O}(\omega_{\eta})$ (independent of the Reynolds number) indicating the curvature (or depth) of the potential well of vortices. 
Furthermore, the macroscopic average particle flux \textcolor{black}{(referring to the bubble velocity oscillation amplitude, $v_0/v_{\text{free}}$, which is a response parameter under the effect of the external forcing)} displays nonlinear trends for increasing $a_\text{e}$ (while $\omega$ is fixed), whose changing rate (slope) indicates the overall \textcolor{black}{mobility} properties of the system. By phenomenological classification, the collective particle motion resembles the behaviour of a ``diode'', an ``anti-diode'' and that of a ``pure resistor''.

We introduce a perspective by likening particle behavior to that of a forced damped oscillator. Based on this analogy, a superposition model is created capable of providing a good prediction of the collective particle behavior and the boundaries between different observed regimes and modes.
This model uncovers the ratio of localized to delocalized particles and highlights features of the potential landscape induced by vortices with a connection to the Kolmogorov scales of turbulence, such as well depth and width. The detected size of the vortices (mean radius $\sim 4\eta$) matches closely those obtained through the identification and statistical analysis of vortex structures in a variety of turbulent ﬂows and for a wide range of Reynolds numbers such as HIT\cite{ghira2022characteristics}, channel ﬂows \cite{kang2007dynamics}, mixing layers \cite{tanahashi2001appearance}, and jets \cite{da2011intense,ganapathisubramani2008investigation}. 
This is remarkable because it indicates that using a Lagrangian method (Lagrangian characteristics of particles in turbulence) the Eulerian characteristics, that is, the depth (natural frequency $\omega_0$) and width (the mean radius of vortices) of potential wells, among other turbulent properties, can be inferred.

Our superposition model provides the theoretical basis for answering the questions of the residence time of particles within vortices as well as the characteristics of vortex-induced potential well. We offer a simple yet effective way of tuning light particle behavior, allowing to introduce forcing in specific regions, which may serve as a mechanism for creating anisotropy in small scales and may have effects on the effective viscosity of fluid, with potential applications in drag reduction, energy transportation, etc.
Several forcing protocols to manipulate the behavior of light particles can be implemented in laboratory experiments, including acoustic waves, magnetic excitation, and electric potentials. Simple methods such as shaking the container promise to offer precise control and diverse manipulation capabilities.
Finally, the influence from the forced light particles on the vortex filaments bears the perspective of being able to change the properties of turbulence, which opens up unexplored future directions: integrating additional and more complex physics effects, including particle feedback on turbulence, particle collisions, lift, and buoyancy forces.

\section*{Materials and Methods}
Extended materials and methods can be found in {\color{blue}\textit{SI Appendix}}
\subsection*{Direct numerical simulations of the Navier-Stokes equation}

The data utilized in both the main paper and the Supplementary Information originate from Direct Numerical Simulations (DNS) of statistically equilibrium homogeneous isotropic turbulence (HIT). The simulation domain is a cubic box with a size of $L=2\pi$ and features periodic boundary conditions. We employ a pseudo-spectral method with Adams-Bashforth for temporal advancement and apply a 2/3-dealiasing rule \cite{peyret2002spectral,bec2006acceleration}. The numerical simulation has been validated using various temporal integration schemes, particle interpolation methods, and large-scale forcing techniques \cite{Calzavarinibook}.
We integrated the Lagrangian evolution of the particles following the equation of motion in Eq.~\ref{eq1} of the main paper, where we consider passively advected, dilute suspensions of spherical particles. The particles are modelled as point particles which are characterized in terms of their dynamical properties, i.e., response time and density contrast (with respect to the fluid). In a nondimensional point of view, these properties are the Stokes number and density contrast $\beta$ (details in {\color{blue}\textit{SI Appendix} Section A}).

\subsection*{Analogue forced damped oscillator: superposition model}

The superposition model reveals that the collective behavior of particles conforms to the ideal theoretical solutions ($v_{\text{trap}}$ and $v_{\text{free}}$) through superposition. 
In the main paper, we specified a threshold $\Omega_{\text{trap}}$ above which the particle is considered trapped in the vortex ($\Omega_{\text{trap}} = 2$). It is found that varying $\Omega_{\text{trap}}$ does not affect the observed similarity in trends between $r$ and $\Omega$ (details see {\color{blue}{\textit{SI Appendix}} Section B}).  
The superposition model comprises two main procedures: (1) global optimization over three-free parameters and (2) zero-free parameter model based on the averaged values from procedure (1).
More details about the superposition modelling procedures, the derivation of Eq.~\ref{final} in the main paper, and the calculation of boundaries between different regimes/modes of the phase diagram (equation shown in the legends in Fig.~\ref{FIG5}(d) and (e)) can be found in {\color{blue}\textit{SI Appendix} Section B}.

\subsection*{Calculation of $v_0$ and extended results of $N^3 = 256^3$ considered in the main paper}
The particle oscillates around the vortex core due to the periodic external force, while the vortex moves within the turbulent flow. Performing velocity averaging over all particles cancels out the turbulent fluctuations' effects, This allows us to extract the influence of the external force, revealing the oscillating velocity's amplitude through Fourier transformation (details can be found in {\color{blue}\textit{SI Appendix} Section C}).
The data used in the main paper are from cases of $\beta \rightarrow 3.0$ and $\text{St}=1.0$ with a resolution of $N^3=256^3$. We investigate 1024 different cases of the applied external force, with its amplitude, $a_{\text{e}}$, and angular frequency, $\omega$, varying in a wide range of $a_{\text{e}} \in [1.0, 150.7]$ and $\omega \in [0.2, 7.0]$. Each case involves a total of 16384 particles. 
In the main paper, we present only eight of the explored $a_{\text{e}}$ values. However, a total of thirty-two $a_{\text{e}}$ values have been investigated. More extended results can be found in {\color{blue}\textit{SI Appendix} Fig.~S3}.

\subsection*{Changing Reynolds number, Stokes number and $\beta$}
Firstly, it is found that the detected natural frequency, $\omega_0$, shows little dependence on the Reynolds number, $\text{Re}_L$, which is of order $\omega_0 \sim \mathcal{O}(\omega_{\eta})$ (details see {\color{blue}{\textit{SI Appendix}} Section D}).
Secondly, one may wonder if the resonant phenomenon of light particles is still prominent when St deviates from 1. To address this question, we conducted multiple simulation runs varying St (keeping $\beta = 3$ constant). We found that as St$\leq 10 $, we can still distinguish between the trap (high enstrophy level) and free states (low enstrophy level) of particles, and the resonance still exists.
However, when St$> 10$, the ensemble average enstrophy for particles remains low for various external forces, indicating the particles' free movement dominant by the external force (details are reported in {\color{blue}\textit{SI Appendix} Section E}).
Thirdly, simulations are conducted (all with St$ = 1$ fixed) with $\beta=0.5$ (heavy particle), $\beta=1.0$ (neutral particle), and $\beta=2.0$ (slightly light particle). We show that the behavior of neutral and heavy particles differs significantly from that of light particles, which is beyond the scope of the current study. 
To identify the natural frequency (by the superposition model), the value of $\beta$ can be extended beyond $\beta>2$, indicating a unique resonant phenomenon for slightly or extremely light particles (i.e., $\rho_p/\rho_f <4$) and once again proves the robustness of the superposition model (details in {\color{blue}\textit{SI Appendix} Section F}).

\paragraph*{Data Availability}
All data underlying this work is publicly available on Zenodo at https://doi.org/10.5281/zenodo.13313453 \cite{wang_2024_13313453}

\paragraph*{Acknowledgments}
We are grateful for the support of the Netherlands Organisation for Scientific Research (NWO) for the use of supercomputer facilities (Snellius) under Grant No. 2021.035, 2023.026. This publication is part of the project “Shaping turbulence with smart particles” with project number OCENW.GROOT.2019.031 of the research programme Open Competitie ENW XL which is (partly) financed by the Dutch Research Council (NWO).
% Please include corresponding author, author contribution and author declaration information
%\authorcontributions{%Please provide details of author contributions here. 
% designed the research;  performed experiments;  processed the data;  analyzed and interpreted the data;    wrote the paper.
\paragraph*{Author Contributions} All authors  designed and performed research; Z.W. analyzed data; and all authors wrote the paper.
%}
%\authordeclaration{The authors declare that they have no competing interests.}
%\equalauthors{\textsuperscript{1}A.O.(Author One) and A.T. (Author Two) contributed equally to this work (remove if not applicable).}
%\correspondingauthor{\textsuperscript{1}To whom correspondence should be addressed. E-mail: chaosun@tsinghua.edu.cn.} %{\color{green}To be written.}

% Keywords are not mandatory, but authors are strongly encouraged to provide them. If provided, please include two to five keywords, separated by the pipe symbol, e.g:
%\keywords{ Rayleigh-B\'enard convection $|$ solidification $|$ density anomaly $|$ coupling interaction $|$ ice dynamics} 
%\showacknow{} 
%\bibliographystyle{elsarticle-num}
%\bibliography{pnas-sample}

%}

%%%%%%%%%%%%%%%%%%%%%%%%%%%%%%%%%%%%%
% SUPPLEMENTARY
%\end{document}

\clearpage
\begin{widetext}
\section*{SI: Supplementary Information}
	\section*{Section A: Direct numerical simulations of the Navier-Stokes equation}
\begin{figure}
	\centering
	\includegraphics[width=0.8\textwidth]{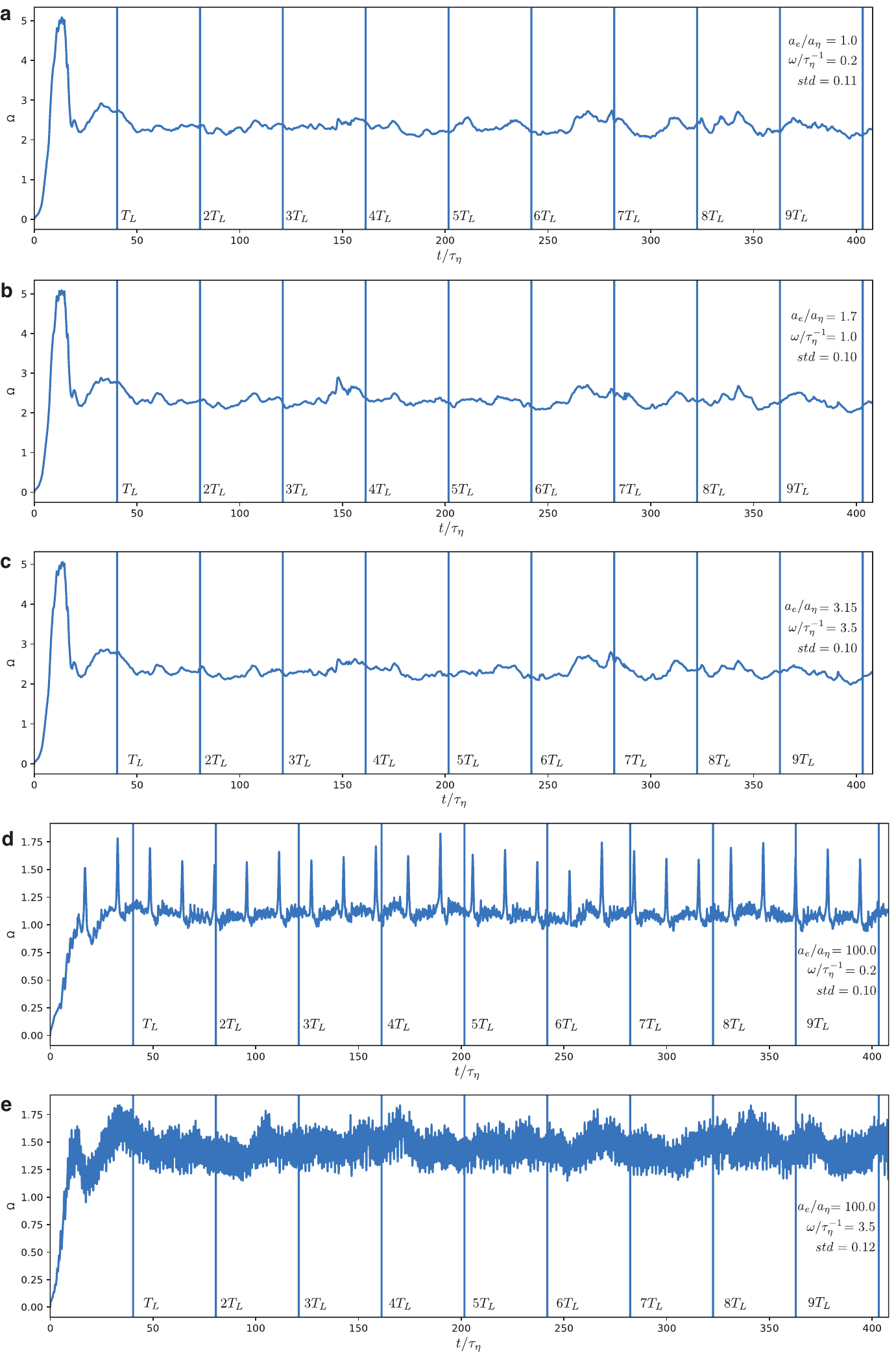}
	\caption{Enstrophy time series.}
	\label{fig0}
\end{figure}

\begin{figure}
	\centering
	\includegraphics[width=\textwidth]{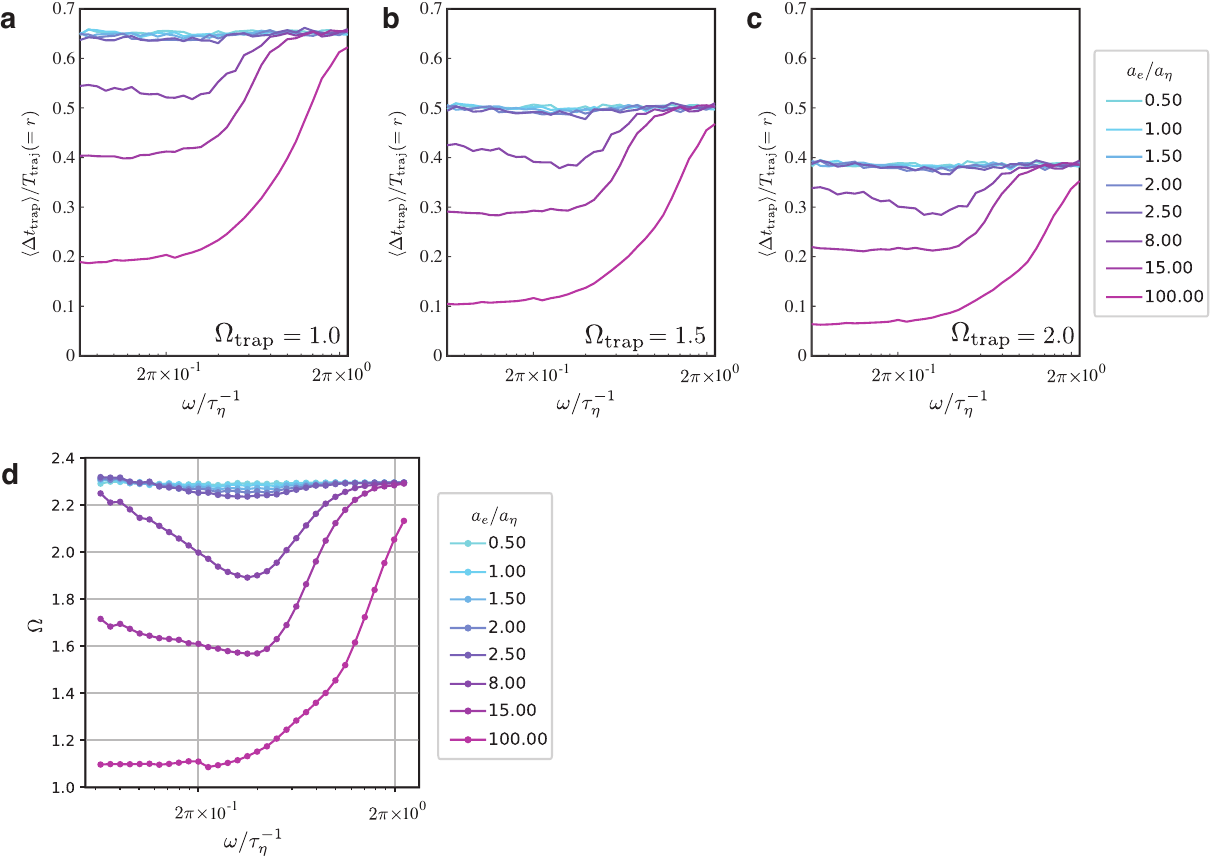}
	\caption{The proportion of localized time under thresholds of $\Omega_{\text{loc}} = 1.0$ (a), $\Omega_{\text{loc}} = 1.5$ (b), and $\Omega_{\text{loc}} = 2.0$ (c), respectively. Comparing these with the enstrophy levels shown in panel (d), it is evident that the similarity in trends between $r$ and $\Delta t_{\text{loc}} / T_{\text{travel}}$ persists, indicating insignificant influence of the threshold value $\Omega_{\text{loc}}$. }
	\label{fig1}
\end{figure}
The data utilized in both the main paper and the Supplementary Information originate from Direct Numerical Simulations (DNS) of statistically equilibrium homogeneous isotropic turbulence (HIT). The simulation domain is a cubic box with a size of $L=2\pi$ and features periodic boundary conditions. We employ a pseudo-spectral method with Adams-Bashforth for temporal advancement and apply a 2/3-dealiasing rule \cite{peyret2002spectralsi,bec2006accelerationsi}. The numerical simulation has been validated using various temporal integration schemes, particle interpolation methods, and large-scale forcing techniques \cite{Calzavarinibooksi}.

We integrated the Lagrangian evolution of the particles following the equation of motion in Eq.1 of the main paper, where we consider passively advected, dilute suspensions of spherical particles. The particles are modelled as point particles which are characterized in terms of their dynamical properties, i.e., response time and density contrast (with respect to the fluid). In a nondimensional perspective, these properties are the Stokes number (St$=\tau_p/\tau_{\eta}$ with $\tau_p$ the particle response time and $\tau_{\eta}$ the Kolmogorov time scale) and $\beta$ ($\beta = \frac{3}{1+2\rho_p / \rho_f}$ with $\rho_p / \rho_f$ representing the density ratio between the particle and the fluid). 

The data used in the main paper are from cases of $\beta \rightarrow 3.0$ ($\rho_p \rightarrow 0$) and $\text{St}=1.0$ with a resolution of $N^3=256^3$. We investigate 1024 different cases of the applied external force, with its amplitude, $a_e$, and angular frequency, $\omega$, varying in a wide range of $a_e \in [1.0, 150.7]$ and $\omega \in [0.2, 7.0]$. Each case involves a total of 16384 particles. We verified convergence by testing different particle numbers, namely 1024, 8192, and 16384, observing that 8192 particles already yield smooth ensemble-averaged results. Increasing the number to 16384 or even further has negligible effects on the results. Considering computational costs and accuracy, we primarily utilize 16384 particles. The corresponding simulation parameters for the DNS used in the main paper are presented in Table.~\ref{simulation_parameters_main}.

Additional simulations are conducted at different resolutions ($N^3 = 128^3, 512^3$) with $\beta \rightarrow 3.0$ and $\text{St}=1.0$ to demonstrate the robustness of our findings, including the revealed mechanism and proposed theoretical model, across Reynolds numbers. The corresponding simulation parameters for the DNS are listed in Table.~\ref{simulation_parameters_changere}.

Furthermore, we vary the characteristics of the particles used, spanning different combinations of their Stokes number ($\text{St} = 0.5, 1.0, 5.0, 10.0, 50.0, 100.0$) and $\beta$ ($\beta = 0.5, 1.0, 2.0, 3.0$). 
When exploring the effect of varying the Stokes number, we keep $\beta = 3$ constant and alter the Stokes number ($\text{St} = 0.5, 1.0, 5.0, 10.0, 50.0, 100.0$).The corresponding simulation parameters are listed in Table.~\ref{simulation_parameters_changest}.
When exploring the effect of varying $\beta$, we fix St$ = 1$ and vary $\beta$ ($\beta = 0.5, 1.0, 2.0, 3.0$). The corresponding simulation parameters are listed in Table.~\ref{simulation_parameters_changebeta}.

We mainly focus on two aspects of the simulation results: 
(1) Temporal evolution features: This pertains to the Lagrangian characteristics of individual particles. For example, Figures 2(f), (h), and (j)) in the main paper illustrate the instantaneous enstrophy at a particle's position during its traversal in the turbulent field.
(2) Collective response of particle Lagrangian features: This refers to the ensemble-averaged behavior or particle-averaged behaviour of particle motion. Here, particles in the system are considered as a collective entity, and their averaged responses are analyzed.
Specifically, the ensemble average of a parameter $f$ of the $i^{th}$ particle at time $t$ is calculated as $\left<f \right>= \left< \left< f_i(t) \right>_{p}\right>_{t} = \frac{1}{n_p(T -t_0)}\int_{t_0}^{T}\sum_{i=1}^{n_p}f_i(t) dt$, where $\left< \cdot \right>$, $\left< \cdot \right>_p$, and $\left< \cdot \right>_t$ denote ensemble average, particle average, and time average, respectively. The time period $t_0$ to $T$ corresponds to the final statistical equilibrium state, and $n_p$ is the total number of particles. An equilibrium state is considered reached after two integral times, $T_L$, after which the standard deviation of the particle-averaged enstrophy time series is less than 0.13, as shown in Fig.~\ref{fig0}. Examples of such analyses (in the main paper) include:
(1) particle average: the proportion of localized time (Fig. 4(e)), resonance curves (Figures 4(g), 5(a), 5(c)), etc.
(2) ensemble average: enstrophy levels (Fig. 4(f)), etc. 

\section*{Section B: Analogue forced damped oscillator: superposition model}
\subsection*{Derivation of the analogue equation of motion for particles in turbulence}
First, we show the details of how to reach the equation of motion for particles in turbulence, which has a similar form to the equation for the forced damped oscillator \cite{den1985mechanicalsi,zill2020advancedsi}.

We start from the Maxey–Riley equation under the conditions of small particle ($\ll$ Kolmogorov length scale $\eta$) and no Fax\'en forces corrections \cite{maxey1983equationsi, mathai2020bubblysi} (non-dimensionalized by the Kolmogorov units of length ($\eta$) and time ($\tau_{\eta}$)), which reads
\begin{equation}
	\begin{split}
		\ddot{\mathbf{x}} _p = \beta \frac{\mathrm{D}\mathbf{u}_f}{\mathrm{D}t} -\frac{1}{\text{St}}( \dot{\mathbf{x}}_p - \mathbf{u}_f) + 
		% (1- \frac{\beta}{3}) 
		a_e \sin(\omega t )\hat{\mathbf{e}}_x,
	\end{split}
	\label{full}
\end{equation}
where $\mathbf{u}_f$ is the flow velocity at the position of the particle, and $\mathbf{x}_p$ is the particle position. The terms $\dot{\mathbf{x}}_p$ and $\ddot{\mathbf{x}}_p$ are the instantaneous particle velocity and acceleration, respectively. Here $\beta = \frac{3}{1+2\rho_p / \rho_f}$ with $\rho_p / \rho_f$ the density ratio between the particle and the fluid, and $\text{St}$ is the Stokes number of the particle. 

From Eq.~\ref{full}, we consider the particle motion in the $x$ direction:
\begin{equation}
	\begin{split}
		\ddot{x}_p = \beta \frac{\mathrm{D} {u}_{fx}}{\mathrm{D}t}  -\frac{1}{\text{St}} (\dot{{x}}_p -u_{fx}) + a_e \sin(\omega t),
	\end{split}
	\label{xdirection}
\end{equation}

Taking the particle average of Eq.~\ref{xdirection}, we get
\begin{equation}
	\begin{split}
		\left<\ddot{x}_p\right>_p = \beta \left<\frac{\mathrm{D} {u}_{fx}}{\mathrm{D}t}\right>_p  -\frac{1}{\text{St}} \left(\left<\dot{ {x}}_p\right>_p -\left<u_{fx}\right>_p \right) + a_e \sin(\omega t ),
	\end{split}
	\label{par_avg}
\end{equation}

Previous works \cite{jimenez1993structuresi, jimenez1998characteristicssi, kang2007dynamicssi, ghira2022characteristicssi} have reported that the small-scale vortex structures can be represented by a classical Burgers vortex model. This model implies a harmonic pressure potential with respect to the position, and we assume the vortex filaments are axisymmetric, so the pressure can be expressed as $P = P_0 + \alpha|\mathbf{x}_p-\mathbf{x}_{vf}|^2
$ with $\alpha$ being the scaling factor and $|\mathbf{x}_p-\mathbf{x}_{vf}|$ being the magnitude of the relative displacement of particle with respect to the vortex core (the vector $(\mathbf{x}_p-\mathbf{x}_{vf})$ is proportional to the core of the vortex). Here, the vector $\mathbf{x}_{vf}=(x_{vf}, y_{vf}, z_{vf})$ is the coordinate of the point located at the vortex core.
The potential well induced force can be calculated by the negative gradient of the potential energy with respect to the position, i.e., $-\nabla P = -(\frac{\partial P}{\partial x}, \frac{\partial P}{\partial y}, \frac{\partial P}{\partial z})$. The potential well induced force in the x direction, i.e., $\frac{\partial P}{\partial x} = 2\alpha(x_p - x_{vf})$, and $\left<x_{vf}\right>_p = 0$ (the origin of the coordinate system is at the center of the cubic box domain). So we can simplify the term of the net effect of added mass and pressure, i.e., $\beta \left< \frac{\mathrm{D}u_{fx}}{\mathrm{D}t} \right>_p$, as  
\begin{equation}
	\begin{split}
		\beta \left< \frac{\mathrm{D}u_{fx}}{\mathrm{D}t} \right>_p &= -\beta \left<\frac{\partial P}{\partial x}\right>_p\\
		&= -2\alpha \beta \left<x_{p} - x_{vf}\right>_p\\
		&= -2\alpha \beta \left<x_{p}\right>_p\\
		&\triangleq -\omega_0^2 \left<x_{p}\right>_p
	\end{split}
	\label{burgers}
\end{equation}  
with $\omega_0 = 2\alpha \beta$ being the averaged natural frequency of the vortex filaments.

Next we consider the term $\left< u_{fx} \right>_p$, which is the Eulerian flow velocity at the position of the particle in Eq.~\ref{par_avg}. The particle averaging leads to $\left< u_{fx} \right>_p = 0$ due to the isotropy of the background turbulence. The particle averaging thus acts as a mechanism to disentangle the collective motion in response to the external force from the local turbulent fluctuations.
Based on the aforementioned arguments and emulating the form of the equation for the forced damped oscillator, we can reorganize Eq.~\ref{par_avg} as,
\begin{equation}
	\begin{split}
		\left<\ddot{x}_p\right>_p = -\omega_0^2 \left<{x}_p\right>_p -\gamma \left<\dot{ {x}}_p\right>_p  + a_e \sin(\omega t ),
	\end{split}
	\label{final}
\end{equation}
with $\gamma = 1/{\text{St}}$ relating the damping force in turbulence. 

Now it can be seen that Eq.~\ref{final} has the same form as that of the forced damped oscillator with natural frequency $\omega_0$, damping $\gamma$, forcing frequency $\omega$ and forcing amplitude $a_e$ \cite{den1985mechanicalsi,zill2020advancedsi}.

\subsection*{The superposition modelling}
To quantify the fraction of time spent inside or outside vortex filaments, we delve deeper into the enstrophy time series. We define a localized interval, $\delta t_{\text{loc}}$, during the light particle trajectory, marked by the condition of enstrophy surpassing a threshold $\Omega_{\text{loc}}$ (dashed line in Fig.~4(a) of the main paper).

% otherwise the particle is free from the vortex either because of the vortex's finite lifetime or because of the strongly dominant external force $\mathbf{f}_{E}$. 
We can categorize the particle motion into two ideal states. As shown in Fig.~4(b) of the main paper, the first state resembles the behavior of a forced damped oscillator (under the influence of $f_D$, $f_P$, and $f_E$), whcih we refer to ``trap'' state.
% , described by control equation Eq.~\ref{final}, where the particles are trapped within dynamic potential wells created by the vortices. In this scenario, due to the fact that the particle motion is dynamically localized in the vortex and the collective mobility of particles is low. 
The velocity oscillation amplitude is $v_{\text{trap}} = a_{\text{e}}\omega/\sqrt{\gamma_{\text{trap}}^2\omega^2+(\omega_0^2 - \omega^2)^2}$, with $\gamma_{\text{trap}}$ being the damping prefactor related to the vortex confinement.
% The potential wells in the turbulent flow exhibit dynamic characteristics, indicating a form of dynamic potential well. 
The second state resembles motion driven by external forcing and dragged by friction force (i.e., under the influence of ${f}_D$ and ${f}_E$, see Fig.~4(c) of the main paper)
% , governed by control equation Eq.~\ref{forced_damped_eqn}. In this state, 
where the particle motion is free from the vortices and freely travelling in the turbulent flow. The particle motion is only minimally influenced by the vortices and their collective mobility is high. 
The amplitude of the velocity oscillation is $v_{\text{free}} = a_{\text{e}}/\sqrt{\gamma_{\text{free}}^2+\omega^2}$, with $\gamma_{\text{free}} =1/\text{St}$ the free damping prefactor. 
Here $\gamma_{\text{trap}}$ and $\gamma_{\text{free}}$ are different in that the particles are subject to varying degrees of constraint by the vortex, leading to differences in the drag forces exerted by the fluid. With this knowledge in mind, Fig.~4(a) of the main paper indicates that the real particle motion is \textit{nonideal} which is a combination of the ``trap'' state (orange shaded range) and the ``free'' state (blue shaded range). 
% These states alternate in the trajectories of particle motion, emphasizing the dynamic nature of the system.
Therefore, the collective amplitude of the velocity oscillation, $v_0$, can be modelled as the superposition of the two ideal states: $v_0^s = rv_{\text{trap}}+(1-r)v_{\text{free}}$. The superposition ratio $r$ bears the physical meaning of the fraction of localized time, which can be calculated by $r=\left<\Delta t_{\text{loc}}\right>_p/T_{\text{traj}}$, where $\Delta t_{\text{loc}} = \sum{\delta t_{\text{loc}}}$ is the proportion of the total localized time interval and $T_{\text{traj}}$ represents the total particle trajectory duration.
Fig.~e(d) of the main paper illustrates the superposed curve (black) resulting from the combination of two velocity amplitude solutions in the ideal states, $v_{\text{trap}}$ (orange dashed line) and $v_{\text{free}}$ (blue dashed line).
% One arises from the fully localized particle (dashed black line). The other originates from the fully delocalized state , $v_{\text{free}}$, indicating the ``free'' state (solid black line). The superposition velocity amplitude, $v_0^s$ (represented by the red line in Fig.~\ref{FIG4}(c)), is then derived by employing the fraction of localized time $r$ as the superposition ratio between $v_{\text{trap}}$ and $v_{\text{free}}$, defined as $v_0^s = rv_{\text{trap}}+(1-r)v_{\text{free}}$. 
The superposition ratio $r$ (the fraction of localized time) as a function of $a_{\text{e}}$ and $\omega$ is plotted in Fig.~4(e) of the main paper ($\Omega_{\text{loc}}$ is choosen as 2). 

Physically, the superposition ratio $r$, when calculating the average enstrophy (ensemble average) experienced by the particle during its motion, measures the relative weight of high enstrophy values. 

% To further explore the relationship between $r$ and the average enstrophy ($\Omega = \left< \Omega(\mathbf{x}_p, t)\right>$, we plot $\Omega$ as a function of $a_{\text{e}}$ and $\omega$ (Fig.~\ref{FIG4}(c)). 
% Here $\left< \cdot \right>$ denotes the ensemble average at the final statistically equilibrium state, and an equilibrium state is considered to be reached after four integral time, $T_L$.
Comparing Fig.~4(e) of the main paper with the enstrophy plot (Fig.~4(f) of the main paper), very similar trends can be observed.
There are two recognizable plateaus: (1) for smaller $a_{\text{e}}$, $\Omega$ remains at a high level (denoted by $\Omega_i$ corresponding to the ``trap'' state due to the less mobility of particle motion with respect to the vortex) no matter how $\omega$ changes. This is because, for weak $\mathbf{f}_{E}$, light particles remain trapped within the vortex filament, where enstrophy remains high. Consequently, the superposition ratio (fraction of localized time) $r$ remains high and relatively constant, although it does not reach unity due to the release of particles caused by the finite lifetime of the vortex.
(2) for large $a_{\text{e}}$, $\Omega$ remains at a low level (denoted by $\Omega_c$ corresponding to the ``free'' state where particle shows free travelling) when $\omega$ is small.
This is because, under strong $\mathbf{f}_{E}$, the external force only significantly influences particle motion when $\omega$ is small enough to allow sufficient time for its effect. Otherwise, particles may not perceive the external force and display locally high frequency oscillation. Correspondingly, the superposition ratio (fraction of localized time) $r$ remains low and relatively constant.

% Moreover, changing the threshold value, $\Omega_{\text{loc}}$, does not affect the observed similarity in the trends of $r$ and $\Omega$ as $a_{\text{e}}$ and $\omega$ change. For example, selecting $\Omega_{\text{loc}}=1.0$ or $1.5$ the similarity persists (see {\color{blue}\textit{SI Appendix} Section B, Fig.~S2}).
% This consistency suggests that the superposition ratio (fraction of localized time) $r$ can be predicted by the ensemble average of $\Omega$, which is a useful aspect for the theoretical modelling.

In the main paper, we specified a threshold $\Omega_{\text{loc}}$ above which the particle is considered trapped in the vortex ($\Omega_{\text{loc}} = 2$). Here, in the Supplementary Information, we demonstrate that varying the threshold value, $\Omega_{\text{loc}}$, does not affect the observed similarity in trends between $r$ and $\Omega$ (as shown in Fig. \ref{fig1}). Figs. \ref{fig1}(a), (b), and (c) depict the proportion of localized time under thresholds of $\Omega_{\text{loc}} = 1.0$, $\Omega_{\text{loc}} = 1.5$, and $\Omega_{\text{loc}} = 2.0$, respectively. Comparing these with the enstrophy levels shown in Fig. \ref{fig1}(d), it is evident that the similarity in trends between $r$ and $\Delta t_{\text{loc}} / T_{\text{travel}}$ persists, indicating insignificant influence of the threshold value $\Omega_{\text{loc}}$. 

% The time evolution of $\Omega$, once again, underscores that the particle motion is a superposition of the ``trap'' and of the ``free'' states. 
% These states alternate in the trajectories of particle motion, emphasizing the dynamic nature of the system.

% From a more quantitative standpoint, as depicted in Fig.~\ref{FIG4}(d), the sketch illustrates the superposed curve (red) resulting from the combination of two velocity amplitude solutions. One arises from the fully localized particle (under the influence of $f_D$, $f_P$, and $f_E$), $v_{\text{trap}}$ (dashed black line). The other originates from the fully delocalized state (under the influence of $\mathbf{f}_D$ and $\mathbf{f}_E$), $v_{\text{free}}$, indicating the ``free'' state (solid black line). The superposition velocity amplitude, $v_0^s$ (represented by the red line in Fig.~\ref{FIG4}(c)), is then derived by employing the fraction of localized time $r$ as the superposition ratio between $v_{\text{trap}}$ and $v_{\text{free}}$, defined as $v_0^s = rv_{\text{trap}}+(1-r)v_{\text{free}}$. 

We map the superposition ratio (fraction of localized time) $r$ from the enstrophy value through a linear transformation, specifically, $r(a_{\text{e}}, \omega) = k \tilde{\Omega}(a_{\text{e}}, \omega)$ where $k$ is the scaling factor. Here, $\tilde{\Omega} = (\Omega-\Omega_c)/(\Omega_i-\Omega_c) \in [0,1]$ represents the normalized enstrophy and $r\in [0,1]$, with $\Omega_i$ and $\Omega_c$ denoting the enstrophy levels in the ``trap'' state (i.e., particles trapped by the vortex) and in the ``free'' state (i.e., particles freely travelling), respectively. Importantly, they only depend on $Re_{L}$ (see Fig.~\ref{fig3} and Fig.~\ref{fig4}). 
Our next task is to utilize the superposition model to predict the behavior of the particle. 
% The three free parameters in the model are $k$, $\omega_0$, and $\gamma_{\text{trap}}$, which can be fitted by the simulation results. The modelling procedures are as follows. First, we explore the global optimal solution for the three free parameters for different $a_{\text{e}}$, i.e., $k(a_{\text{e}})$, $\omega_0(a_{\text{e}})$, and $\gamma_{\text{trap}}(a_{\text{e}})$ (plots can be found in {\color{blue}\textit{SI Appendix} Fig.~S2}). It is found that within the range of $a_{\text{e}}<5$, where the natural frequency $\omega_0$ can still be detected, the three parameters, $k(a_{\text{e}})$, $\omega_0(a_{\text{e}})$, and $\gamma_{\text{trap}}(a_{\text{e}})$, remain constant. Based on this fact, we calculate the average, $k$, $\omega_0$, and $\gamma_{\text{trap}}$, of the three free parameters in the range of $a_{\text{e}}<5$. By substituting these averaged values, i.e., $k\approx 0.42$, $\omega_0 \approx 1.47$, and $\gamma_{\text{trap}}\approx 1.62$, ($k \approx 0.46 \pm 0.4$, $\omega_0 \approx 1.45 \pm 0.11$, and $\gamma_{\text{trap}} \approx 1.65 \pm 0.08$ as Reynolds number changes, details in {\color{blue}\textit{SI Appendix} Fig.~S3-S4}), the superposition model degenerates into one with no free parameter.} 

The superposition model comprises two main procedures. 

Initially, we note that it includes three free parameters: $\gamma_{\text{trap}}$, $\omega_0$, and $k$. By employing the least squares method to fit the simulation results with the exact form of the equation $v_0^s = rv_{\text{trap}}+(1-r)v_{\text{free}}$, we obtain the globally optimal values for these three parameters, denoted as $k(a_e)$, $\omega_0(a_e)$, and $\gamma_{\text{trap}}(a_e)$, all of which are functions of $a_e$ (as shown in Fig.~\ref{fig2}(c-e)). It can be observed from the figure that when $a_e<5$, the parameters $k(a_e)$, $\omega_0(a_e)$, and $\gamma_{\text{trap}}(a_e)$ remain nearly unchanged (dark symbols in Fig.~\ref{fig2}(c-e)). Based on this observation, we calculate the averages $\gamma_{\text{trap}}$, $\omega_0$, and $k$ of the three free parameters within the range of $a_e<5$, represented by the red-dotted line ($\gamma_{\text{trap}}$) in Fig.~\ref{fig2}(c), the red-dashed line ($\omega_0$) in Fig.~\ref{fig2}(d), and the red-dashed line ($k$) in Fig.~\ref{fig2}(e), respectively.

Second, by substituting these averaged values, i.e., $k\approx 0.42$, $\omega_0 \approx 1.47$, and $\gamma_{\text{trap}}\approx 1.62$, the superposition model degenerates into one with no free parameter.

Plotting the theoretical results (black dashed lines in Fig.~\ref{fig2}(d), and black dashed lines in Fig.~4(g) of the main paper) and the numerical simulation results (symbols in Fig.~\ref{fig2}(d), and symbols in Fig.~4(g) of the main paper), an excellent agreement can be observed. Here $v_0$ is the particles averaged velocity oscillation amplitude which is extracted by performing a Fourier transform on the periodically oscillating particle-averaged velocity signal (details in {\color{blue}Section C})). 

One might question why we can utilize the averaged values of $k$, $\omega_0$, and $\gamma_{\text{trap}}$ for $a_{\text{e}} < 5$ to predict particle behavior across a broad range of $a_{\text{e}}$. The reason is that as $a_{\text{e}}$ increases, particles tend to behave similarly to the free-traveling mode, where $r$ in $v_0^s = rv_{\text{trap}}+(1-r)v_{\text{free}}$ approaches zero, implying $v_0^s \approx v_{\text{free}}$. It's important to note that $v_{\text{free}} = a_{\text{e}} / \sqrt{(1/\text{St}^2 + \omega^2)}$ is independent of $\omega_0$ and $\gamma_{\text{trap}}$.

% Moreover, it's noteworthy that as the Reynolds number varies, these three parameters—$k$, $\omega_0$, and $\gamma_{\text{trap}}$—remain relatively stable ($k \approx 0.46 \pm 0.4$, $\omega_0 \approx 1.45 \pm 0.11$, and $\gamma_{\text{trap}} \approx 1.65 \pm 0.08$) (detailed results can be found in {\color{blue}\textit{SI Appendix} Fig.~S4-S5}), indicating the robustness of the superposition model.
Moreover, it's noteworthy that as the Reynolds number varies, these three parameters—$k$, $\omega_0$, and $\gamma_{\text{trap}}$—remain relatively stable ($k \approx 0.46 \pm 0.4$, $\omega_0 \approx 1.45 \pm 0.11$, and $\gamma_{\text{trap}} \approx 1.65 \pm 0.08$). Detailed results can be found in Fig.~\ref{fig3} and Fig.~\ref{fig4}, indicating the robustness of the superposition model.

% The superposition model reveals that the collective behavior of particles conforms to the ideal theoretical solutions through superposition. The ideal theoretical solutions include: (1) $v_{\text{trap}}$, the oscillation amplitude of the fully localized particles (under the influence of $f_D$, $f_P$, and $f_E$); (2) $v_{\text{free}}$, which is the oscillation amplitude of the fully delocalized state (under the influence of ${f}_D$ and ${f}_E$). The superposition velocity amplitude, $v_0^s$, is then derived employing a prefactor $r$ as the superposition ratio between $v_{\text{trap}}$ and $v_{\text{free}}$, defined as $v_0^s = rv_{\text{trap}}+(1-r)v_{\text{free}}$. 
% $r$ has exact physical meaning of the proportion of the localized time, i.e., $r = \Delta t_{\text{loc}}/T_{\text{travel}}$. 

\subsection*{Prediction of boundaries between different regimes/modes of the phase diagram}
% First, we show the resonance phenomenon of the collective motion of particles. 
% % Note that hereafter we discuss the ensemble average (macroscopic average response) instead of focusing on the time evolution. 
% Fig.~\ref{FIG5}(a) shows the normalized velocity oscillation amplitude, $v_0/v_{\text{free}}$, as a function of the forcing frequency $\omega$ of the external force.  
By the superposition model, the predicted ensemble velocity oscillation amplitude normalized by $v_{\text{free}}$ can be written as,
\begin{equation}
\begin{split}
	v_0^s/v_{\text{free}} &= rv_{\text{trap}}/v_{\text{free}} + (1-r),\\
	&=r \left[ \sqrt{\frac{\gamma_{\text{free}}^2\omega^2 + \omega^4}{\gamma_{\text{trap}}^2 \omega^2 +(\omega_0^2-\omega^2)^2}} -1\right] +1.
\end{split}
\label{vi_div_vc}
\end{equation}
We have $v_0^s/v_{\text{free}} = 1$ under the condition of either $r=0$ or $\sqrt{(\gamma_{\text{free}}^2\omega^2 + \omega^4)/(\gamma_{\text{trap}}^2 \omega^2 +(\omega_0^2-\omega^2)^2})=1$. 
On the one hand, when $a_{\text{e}}\sim\mathcal{O}(10^2)$ we have $r\rightarrow 0$ (see Fig.~4(e) of the main paper). On the other hand, when $\omega \gg\omega_{\eta}$ the term in brackets of Eq.~\ref{vi_div_vc} tends towards 0 because $\omega_0 \approx 1.5 \sim\mathcal{O}(\omega_{\eta})$ (dotted line in Fig.~5(a) of the main paper which is detected by the superposition model, recall that $\omega_{\eta}=2\pi/\tau_{\eta} = 2\pi$). Both conditions leads to the same particle response, i.e., $v_0^s/v_{\text{free}} \approx 1$. However, these two conditions have different physical meanings: (1) $a_{\text{e}} \gg \mathcal{O}(10^2)$ indicates that the external force is strong enough to drive particles to escape from the vortex constraint, displaying a prominent periodic oscillation mode in a wide spatial range; (2) $\omega \gg\omega_{\eta}$ indicates that the particle position oscillation amplitude satisfies $x_0 \propto a_{\text{e}}/\omega^2$ (see Eq.~5 of the main paper) independent of $\omega_0$. 

The condition of $v_0/v_{\text{free}} > 1$ represents resonance coming into effect. The resonant frequency, $\hat{\omega}_0$ (solid vertical line in Fig.~5(a) of the main paper), is slightly larger than the natural frequency, $\omega_0$. 
The resonant frequency, $\hat{\omega}_0$, is slightly modulated by the damping, which is expressed as 
\begin{equation}
\begin{split}
	\hat{\omega} = \sqrt{\frac{\omega_0^4 + \sqrt{\omega_0^4 (\gamma_{\text{free}}^2 - \gamma_{\text{free}} \gamma_{\text{trap}} + \omega_0^2) (\gamma_{\text{free}} (\gamma_{\text{free}} + \gamma_{\text{trap}}) + \omega_0^2)}}{\gamma_{\text{free}}^2 - \gamma_{\text{trap}}^2 + 2 \omega_0^2}}.
\end{split}
\end{equation}

While the condition of $v_0/v_{\text{free}} < 1$ manifests the constraints imposed by the vortex on the particle motion. Fig.~5(a) of the main paper tells us that as $a_{\text{e}}$ is chosen at low and intermediate levels ($a_{\text{e}}<5 a_{\eta}$), the resonance occurs from which $\omega_0$ (and thus $\hat{\omega}_0$) can be detected (blue and light red curves). But this is not the case for the situation of very large $a_{\text{e}}$ where the particle is dominated by the external force and no resonance exists (dark red symbols in Fig.~5(a) of the main paper). 

The detected natural frequency, $\omega_0$ (dashed line in Fig.~5(a) of the mian paper), shows little dependence on the Reynolds number, $\text{Re}_L$, as shown by symbols with error bar in Fig.~5(b) of the mian paper. The green dashed line, red dotted line, and blue solid line shows scaling characteristics (obtained by scaling analysis \cite{frisch1995turbulencesi}) of $\omega_0 \sim 1/T_{L} \propto \text{Re}_{\text{L}}^{-1/2}$, $\omega_0 \sim 1/\tau_{\lambda} \propto \text{Re}_{\text{L}}^{-1/6}$, and $\omega_0 \sim 1/\tau_{\eta} \propto \text{Re}_{\text{L}}^{0}$ respectively. The detected natural frequency is of order $\omega_0 \sim \mathcal{O}(\omega_{\eta})$, which indicates the depth of the potential well felt by the particle.

From a more physical point of view, the normalized particle velocity oscillation amplitude, $v_0/v_{\text{free}}$ can be regarded as a macroscopic average particle flux: it accounts for the collective mobility of the particles. Different nonlinear responses can be identified upon gradually increasing $a_{\text{e}}$ (for a fix $\omega$).
The amplitude of the external force, $a_{\text{e}}$, acts as a driving force (analogous to voltage), and the macroscopic average particle flux, $v_0/v_{\text{free}}$, serves as a response (analogous to current). As the driving force increases, the changing rate (slope) of the response indicates the overall impedance properties of the system. 
It is found that for different $\omega$ the macroscopic average particle flux $v_0/v_{\text{free}}$ displays different trends as increasing $a_{\text{e}}$, which are, by phenomenological classification, ``diode'' (response switch-on), ``anti-diode'' (response shut-down), and ``pure resister'' (response unchanged).

% Localization-delocalization transition for light particles has been identified upon increasing $a_{\text{e}}$ (for a fix $\omega$). 
% As shown in Fig.~\ref{FIG5}(c) is the normalized particle velocity oscillation amplitude, $v_0/v_{\text{free}}$, as varying $a_{\text{e}}$.
% For different $\omega$, the macroscopic average particle flux $v_0/v_{\text{free}}$ displays different {\color{red}trends} as increasing $a_{\text{e}}$.
For smaller $\omega$, the response parameter $v_0/v_{\text{free}}$ exhibits a ``diode'' alike regime, where $v_0/v_{\text{free}}$ firstly remains at a smaller level and almost unchanged, as $a_{\text{e}}$ increases $v_0/v_{\text{free}}$ experiences a ``switch-on'' transitional (abrupt increase) and reaches a higher level at large $a_{\text{e}}$. The feature of this transition is highlighted by the sketch in Fig.~5(f) of the main paper. Unlike the real diode, the particles in the diode mode still display some mobility in the smaller $a_{\text{e}}$ range, indicating some ``leaky current'' under weak driving force. This leaky current is induced by the finite lifetime of the vortex which leads to occasionally freely moving particles. This can be verified when the Eulerian flow field is frozen, where the flow field is enforced to stop update and to remain the same as the initial state and thus the vortices always exist, the leaky current drops a lot (see the blue line in the inset Fig.~5(c) of the main paper, note that this ``leaky current'' is not zero due to the finite size effect of the vortex filaments in turbulence).
For intermediate $\omega$, the response parameter $v_0/v_{\text{free}}$ presents an ``anti-diode'' like regime, where as $a_{\text{e}}$ increases the system exhibits a ``shut-down'' transitional and finally arrives at a ``immobilized'' state (whose feature is highlighted by the sketch in Fig.~5(g) in the main paper). 
Note that the immobilized state is not really absence of mobility, instead, there is some leaky current (limited mobility of low level) resulting from the vortex finite lifetime which has been discussed before.
For high $\omega$, the particles enter the purely external force driven state, which is like the pure resistor (whose feature is highlighted by the sketch in Fig.~5(h) of the main paper).
The threshold of $\omega$, $\omega^*$, that control the boundary between the diode and anti-diode mode can be determined utilizing the superposition model. 
As is shown in Fig.~4(b), the superposition ratio (fraction of localized time, or equivalently the fraction of trapped particles) $r$ is monotonically increasing with respect to $a_{\text{e}}$, which means when the coefficient (terms enclosed in square brackets) of $r$ in Eq.~\ref{vi_div_vc} is greater than 0, $ v_0/v_{\text{free}}$ has a positive slope. Therefore $\omega < \omega* =\left(\omega_0^4/(\gamma_{\text{free}}^2 + 2\omega_0^2-\gamma_{\text{trap}}^2) \right)^{1/2}$ gives rise to the diode mode. On the other hand, $v_0^s/v_{\text{free}} = 1$ under the condition of $a_{\text{e}} \gg \mathcal{O}(10^2)$ or $\omega \gg \omega_{\eta}$, $v_0^s/v_{\text{free}} \approx 1$ so the particle enters what is the pure resistor mode (with a constant resistance level). 
The range $\omega^* < \omega \ll \omega_{\eta}$ results in the anti-diode mode. In the limit of $\gamma_{\text{trap}} = \gamma_{\text{free}}$, we have $\omega^* \sim \omega_0/\sqrt{2}$. 

From what we discussed before, it has been shown that the phenomenology of particle behaviors can change drastically, e.g., by modifying the properties of the external force ($a_{\text{e}}$ and $\omega$). A general classification of conditions, under which exact particle behaviors are expected to happen, are (1) specific external force with $a_{\text{e}}$ and $\omega$ is applied; (2) for a fixed $a_{\text{e}}$, $\omega$ increases gradually; (3) for a fixed $\omega$, $a_{\text{e}}$ increases gradually. Detailed answers to these questions are displayed in the phase diagram (see Fig.~5(d)-(h) in the main paper). 

When a specific external force with $a_{\text{e}}$ and $\omega$ is applied to the particle, the particle behavior has three identified regimes depending on $a_{\text{e}}$, which is plotted in the phase diagram colored by different levels of $r$. One can expect that when the particle position oscillates within some threshold the particle can be well trapped in the vortex, i.e., $x_0<x_0^{\text{loc}}$. Similarly, when $x_0>x_0^{\text{deloc}}$ the particle can be regarded as free travelling. A transitional regime is identified when the particle oscillates in the range from $x_0^\text{loc}$ to $x_0^\text{deloc}$. The values of $x_0^\text{loc}\sim 4 \eta$ and  $x_0^\text{deloc}\sim 12 \eta$ are determined by fitting the simulation results (recall that values are in the unit of Kolmogorov length scale, $\eta$). These two constraints pose two boundaries between the localized and transitional regimes as well as the transitional and delocalized regimes, respectively. 
The solid and dashed lines indicate the boundary between different regimes, which are defined by $a_{\text{e}} = x_0^{\text{loc}} \sqrt{\gamma_{\text{trap}}^2 \omega^2 + (\omega_0^2 - \omega^2)^2}$ and $a_{\text{e}} = x_0^{\text{deloc}} \sqrt{\gamma_{\text{trap}}^2 \omega^2 + (\omega_0^2 - \omega^2)^2}$, respectively.
Note that the two thresholds, $x_0^{\text{loc}}$ and $x_0^{\text{deloc}}$, are set because the potential well is not perfectly sharp, and it has a diffusive boundary instead. In fact, the oscillating amplitude of the particle position in the transitional regime indicates that the diffusive boundary of the potential well (induced by the vortex) felt by the light particles in HIT is monodisperse, with a range approximately from $4\eta$ to $12 \eta$. Furthermore, $x_0^\text{loc}$ ($\sim4 \eta$) indicates the upper bound of the oscillating amplitude of the particle position in the localized regime, which implies the mean radius of the vortex in turbulence is approximately $4\eta$, which is comparable to those obtained through the identification and statistical analysis of vortex structures in a variety of turbulent ﬂows and for a wide range of Reynolds numbers \cite{tanahashi2001appearancesi,kang2007dynamicssi, ghira2022characteristicssi,da2011intensesi,ganapathisubramani2008investigationsi}. 

We can also understand the motion of particles from another perspective as shown in Fig.~5(e) of the main paper. The normalized velocity oscillation amplitude, $v_0/v_{\text{free}}$ bears the physical meaning of the mobility of the particle. When $v_0/v_{\text{free}}<1$, the particle is in an immobilized state under the constraints imposed by the vortex and typically drifts with the vortex (blue-colored region in the phase diagram shown in Fig.~5(e) in the main paper). $v_0/v_{\text{free}}>1$ happens because of the resonance (red-colored region in the phase diagram shown in Fig.~5(e) in the main paper). As $v_0/v_{\text{free}}=1$ the particle displays a neutral mode (white-colored region in the phase diagram shown in Fig.~5(e) in the main paper).

Combing the information revealed in the first phase diagram (Fig.~5(d)), we can obtain a more detailed classification and a better understanding of the particle behavior: by choosing $a_{\text{e}}$ one can enforce the particle in a specific regime, i.e., localized, transitional, and delocalized. As the particle in the localized regime, by choosing $\omega$ the particle may land in different modes, i.e., immobilized, resonant, and neutral. Here, these three modes are mostly considered in the localized regime which is because with the confinement of the vortex potential well, the particle behaves more like a forced damped oscillator where prominent resonance occurs (a dark red patch in the phase diagram of Fig.~5(e) of the main paper). While in the transitional regime, the particle escapes the vortex more often leading to less obvious or polluted (by particle free travelling) resonance (a light red patch in the phase diagram of Fig.~5(e) of the main paper). The particle in the delocalized regime is totally in the neutral mode (a white patch in the phase diagram of Fig.~5(e) of the main paper) with no indication of the resonance anymore.
This discussion infers that the natural frequency of the potential well can be detected by choosing a proper $a_{\text{e}}$ within the localized regime (roughly $a_{\text{e}}<10$ but smaller $a_{\text{e}}$, e.g., $a_{\text{e}}<5$, can give a more accurate estimation of $\omega_0$) and increasing $\omega$ gradually, and $\omega_0$ can be predicted by the superposition model.

For a fixed $\omega$ and gradually increasing $a_{\text{e}}$,  the particle displays a transition process following ``diode'' ($\omega<\omega^*$), anti-diode (as $\omega^*<\omega \ll \omega_{\eta}$), or pure resistor ($\omega^* > \omega_{\eta}$) pattern. 

\section*{Section C: Calculation of $v_0$ and extended results of $N^3 = 256^3$ considered in the main paper}
\begin{figure}
	\centering
	\includegraphics[width=\textwidth]{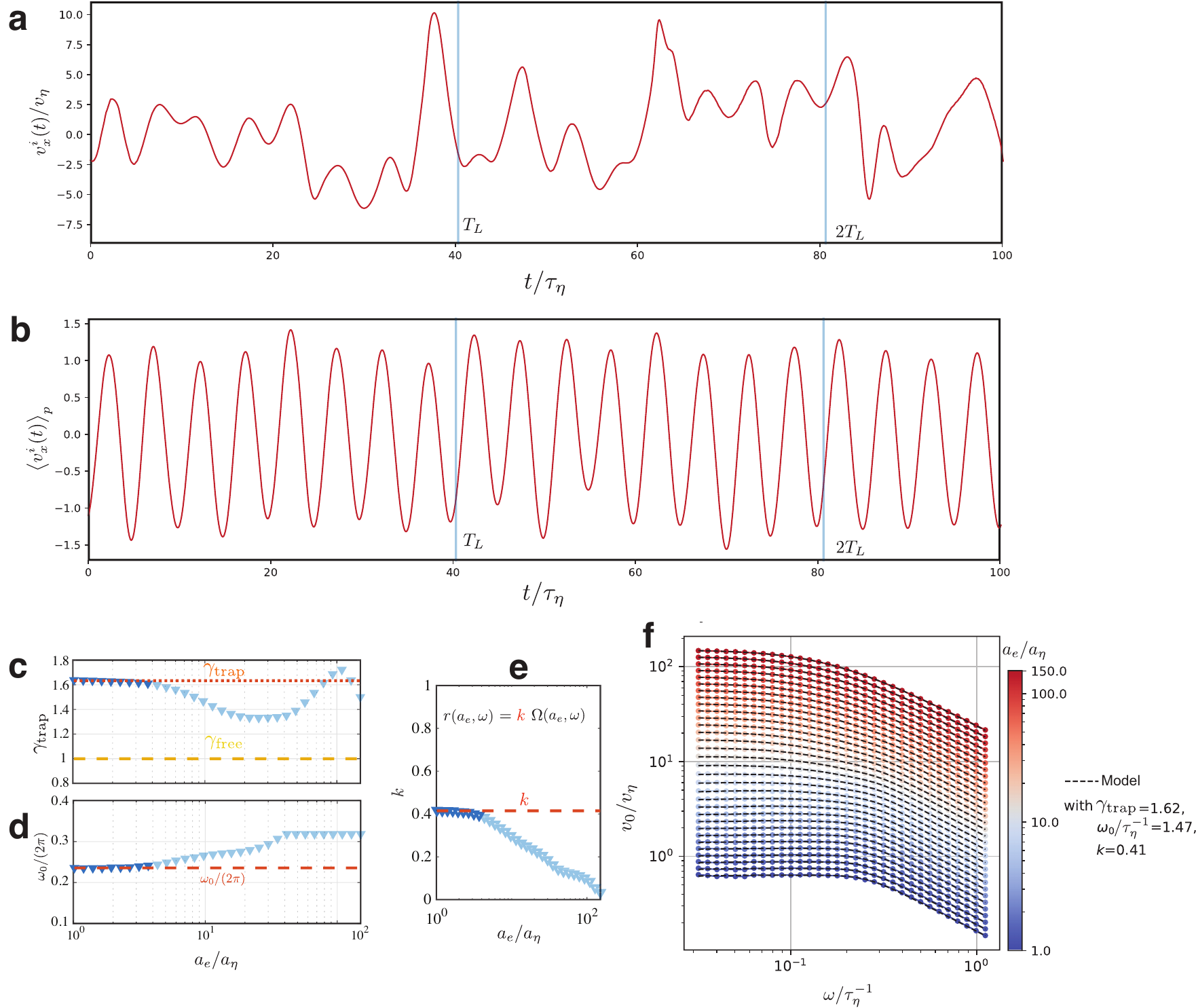}
	\caption{(a) Single particle velocity ($x$ component, denoted as $v_x^i(t)$) time series.  An intensive fluctuations can be observed. The external force is $\omega = 1.25$ and $a_e =  2.0$. (b) Particle averaged velocity, i.e., $\left< v_x^i(t) \right>_p$, time series. An organized periodic oscillation pattern can be observed. The external force is $\omega = 1.25$ and $a_e =  2.0$.
		The globally optimal values for the three free parameters in the superposition model, denoted as $\gamma_{\text{trap}}(a_e)$ (a), $\omega_0(a_e)$ (b), and $k(a_e)$ (c), all of which are functions of $a_e$. It can be observed from the figure that when $a_e<5$, the parameters $k(a_e)$, $\omega_0(a_e)$, and $\gamma_{\text{trap}}(a_e)$ remain nearly unchanged (dark symbols). Based on this observation, we calculate the averages $\gamma_{\text{trap}}$, $\omega_0$, and $k$ 
		within the range of $a_e<5$, represented by the red-dotted line ($\gamma_{\text{trap}}$) in panel (a), the red-dashed line ($\omega_0$) in panel (b), and the red-dashed line ($k$) in panel (c), respectively.
		By substituting these averaged values, i.e., $k\approx 0.42$, $\omega_0 \approx 1.47$, and $\gamma_{\text{trap}}\approx 1.62$, the superposition model degenerates into one with no free parameter.
		Plotting the theoretical results (black dashed lines in panel (d)) and the numerical simulation results (symbols in panel (d)), a good agreement can be observed. }
	\label{fig2}
\end{figure}
\subsection*{Calculation of $v_0$}
The particle oscillates with respect to the vortex core under the influence of the periodic external force, while the vortex itself sweeps around (turbulent fluctuations) within the turbulent flow field. The output of the numerical simulation of the velocity in the $x$-direction for the $i^{th}$ particle (denoted as $v_x^i(t)$) is a combination of these two effects. The time series of $v_x^i(t)$, shown in Fig.~\ref{fig2}(a) (a typical example under the external force of $\omega = 1.25$ and $a_e =  2.0$), displays intensive fluctuations. 
We first average the $x$-component velocity over all the particles, i.e., $\left< v_x^i(t) \right>_p$, by which the effect from the turbulent fluctuations on the particle motion can be cancelled out. The times series of $\left< v_x^i(t) \right>_p$, shown in Fig.~\ref{fig2}(b) (still under the external force of $\omega = 1.25$ and $a_e =  2.0$), exhibits a clear sinusoidal pattern, indicating the effect from the external force has been extracted and the turbulent fluctuations has been cancelled out. The amplitude of the oscillatory velocity (denoted as $v_0$) is obtained by performing a Fourier transform on $\left< v_x^i(t) \right>_p$, corresponding to the intensity at the characteristic frequency. 

\subsection*{Extended results of $N^3 = 256^3$ considered in the main paper}
In the main paper, we present only eight of the explored $a_e$ values. However, a total of thirty-two $a_e$ values have been investigated. The extended results of the comparison between the simulation and the superposition modeling are depicted in Fig. \ref{fig2}(f).

\section*{Section D: Changing Reynolds number}
With the same external force, we change the Reynolds number of the system, and the results are depicted in Fig.~\ref{fig3} and Fig.~\ref{fig4}.
\begin{figure}
	\centering
	\includegraphics[width=\textwidth]{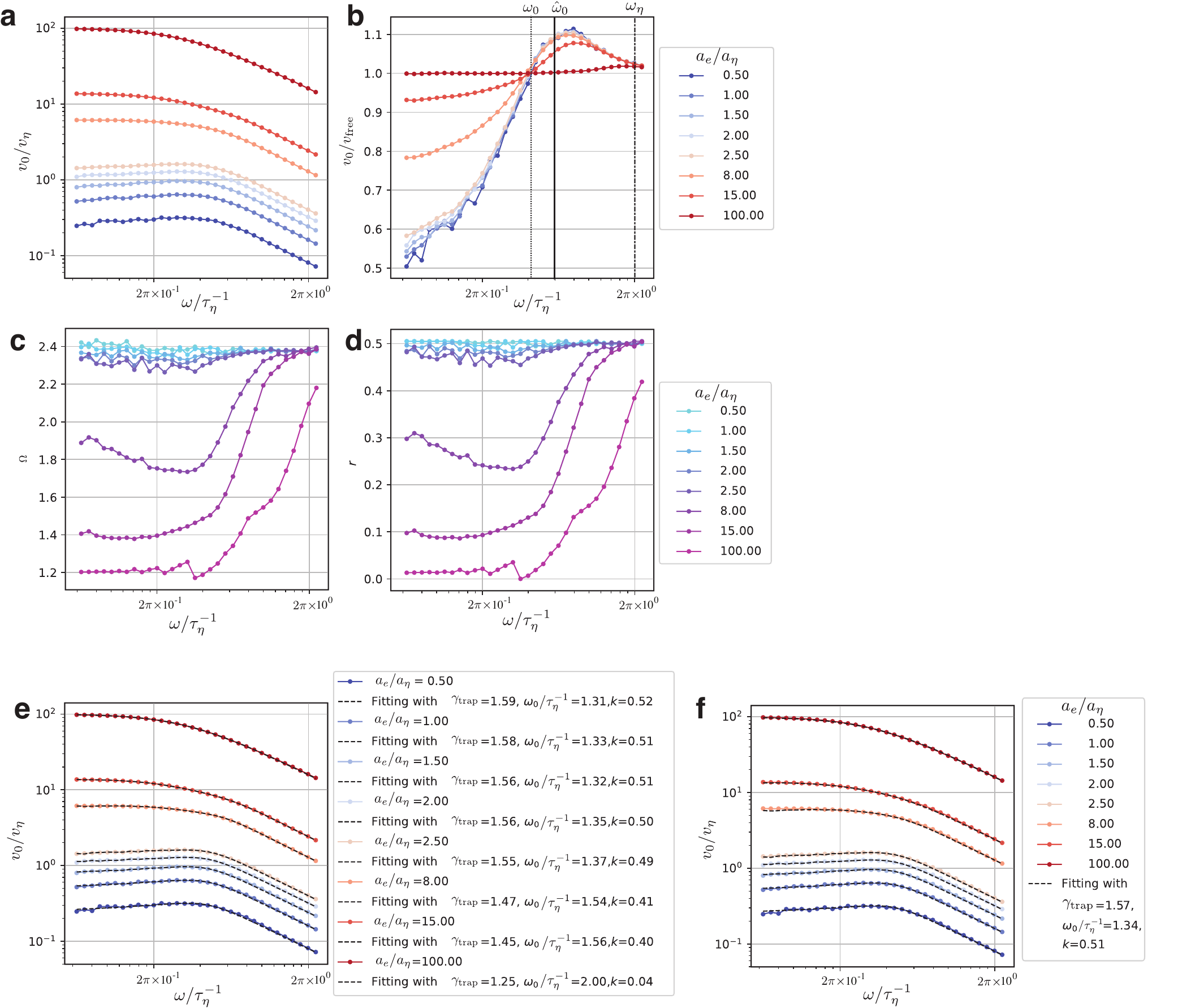}
	\caption{Results of resolution: $N^3 = 128^3$. The input parameters of the simulation are presented in the first row of Table \ref{simulation_parameters_changere}. (a) The original velocity oscillation amplitude $v_0$. (b) The normalized velocity oscillation amplitude $v_0/v_{\text{free}}$ as a function of $a_e$ and $\omega$. (c) The enstrophy level $\Omega$ as a function of $a_e$ and $\omega$. (d) The superposition prefactor $r$ as a function of $a_e$ and $\omega$. (e) The three-parameter superposition modeling results obtained through the first procedure described in Section B. (f) The final superposition modeling results obtained after the second procedure described in Section B.
	}
	\label{fig3}
\end{figure}

\begin{figure}
	\centering
	\includegraphics[width=\textwidth]{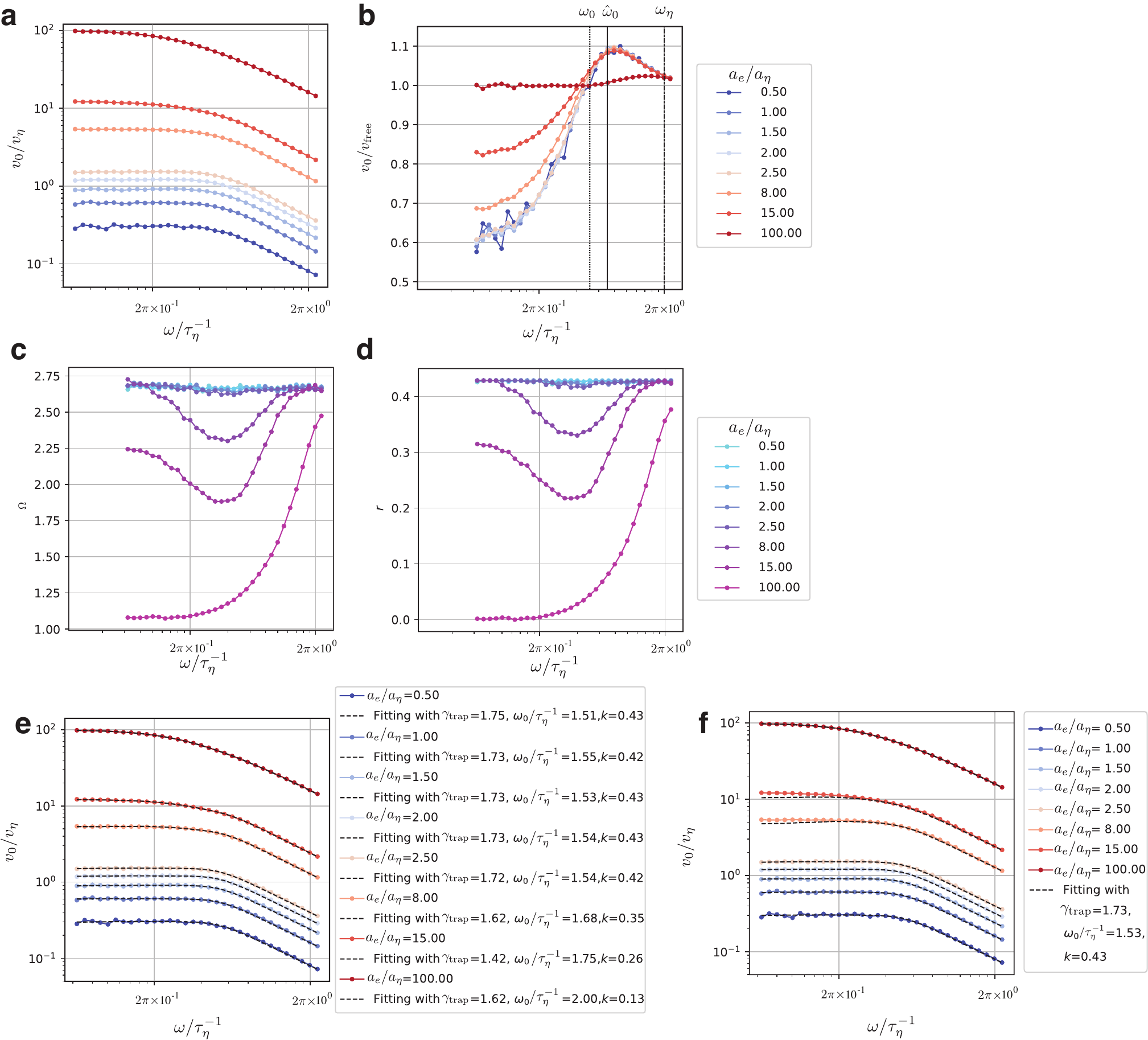}
	\caption{Results of resolution: $N^3 = 512^3$. The input parameters of the simulation are presented in the second row of Table \ref{simulation_parameters_changere}. (a) The original velocity oscillation amplitude $v_0$. (b) The normalized velocity oscillation amplitude $v_0/v_{\text{free}}$ as a function of $a_e$ and $\omega$. (c) The enstrophy level $\Omega$ as a function of $a_e$ and $\omega$. (d) The superposition prefactor $r$ as a function of $a_e$ and $\omega$. (e) The three-parameter superposition modeling results obtained through the first procedure described in Section B. (f) The final superposition modeling results obtained after the second procedure described in Section B.
	}
	\label{fig4}
\end{figure}

\subsection*{Results of a smaller resolution: $N^3 = 128^3$}
The input parameters of the simulation are presented in the first row of Table \ref{simulation_parameters_changere}. Fig.~\ref{fig3} shows the original velocity oscillation amplitude $v_0$ (Fig. \ref{fig3}(a)), the normalized velocity oscillation amplitude $v_0/v_{\text{free}}$ (Fig. \ref{fig3}(b)), the enstrophy level $\Omega$ (Fig. \ref{fig3}(c)), the superposition prefactor $r$ (Fig. \ref{fig3}(d)), the three-parameter superposition modeling results obtained through the first procedure described in Section B (Fig. \ref{fig3}(e)), and the final superposition modeling results obtained after the second procedure described in Section B (Fig. \ref{fig3}(f)).

\subsection*{Results of a higher resolution: $N^3 = 512^3$}
The input parameters of the simulation are displayed in the second row of Table \ref{simulation_parameters_changere}. Specifically, the original velocity oscillation amplitude $ v_0 $ (Fig. \ref{fig4}(a)), the normalized velocity oscillation amplitude $ v_0 / v_{\text{free}} $ (Fig. \ref{fig4}(b)), the enstrophy level $ \Omega $ (Fig. \ref{fig4}(c)), the superposition prefactor $ r $ (Fig. \ref{fig4}(d)), the three-parameter superposition modeling results obtained through the first procedure outlined in Section B (Fig. \ref{fig4}(e)), and the final superposition modeling results obtained after the second procedure described in Section B (Fig. \ref{fig4}(f)) are presented.

\section*{Section E: Changing Stokes number}
\begin{figure}
	\centering
	\includegraphics[width=\textwidth]{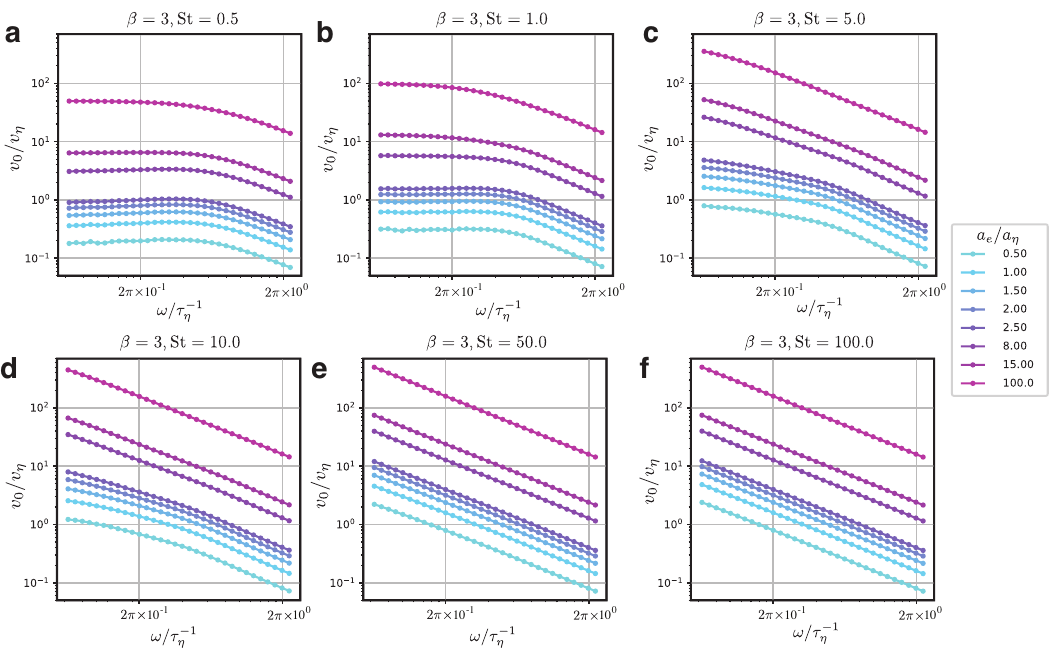}
	\caption{The original velocity oscillation amplitude $v_0$ with different Stokes numbers.}
	\label{fig5}
\end{figure}
\begin{figure}
	\centering
	\includegraphics[width=\textwidth]{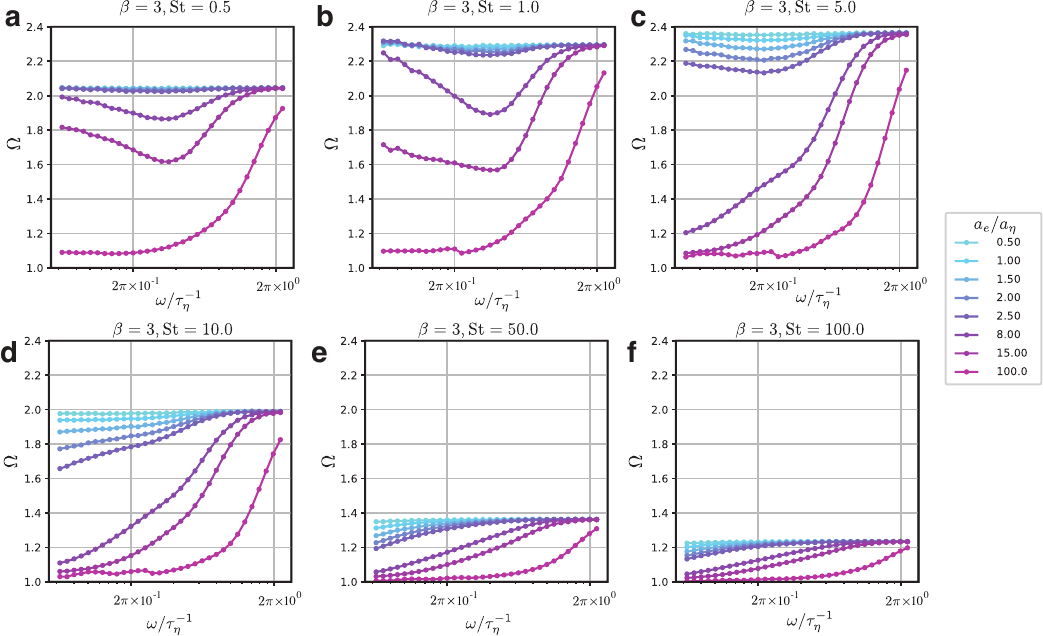}
	\caption{ The enstrophy as a function of $a_e$ and $\omega$ for different Stokes numbers.}
	\label{fig6}
\end{figure}
\begin{figure}
	\centering
	\includegraphics[width=\textwidth]{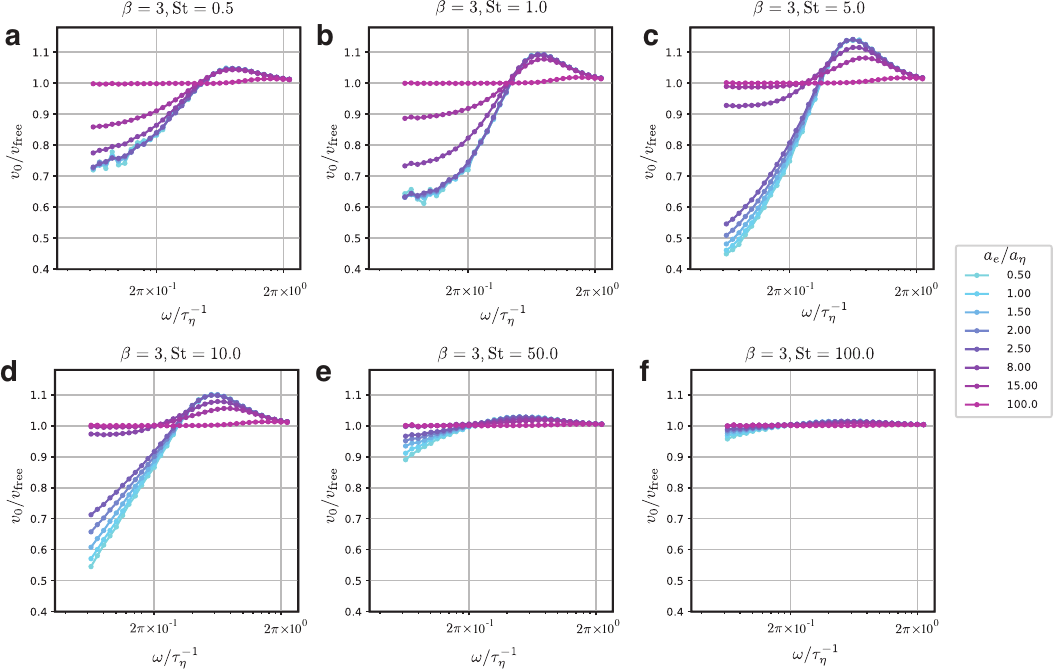}
	\caption{The normalized response $v_0/v_{\text{free}}$ as a function of $a_e$ and $\omega$ for different Stokes numbers.}
	\label{fig7}
\end{figure}
\begin{figure}
	\centering
	\includegraphics[width=0.7\textwidth]{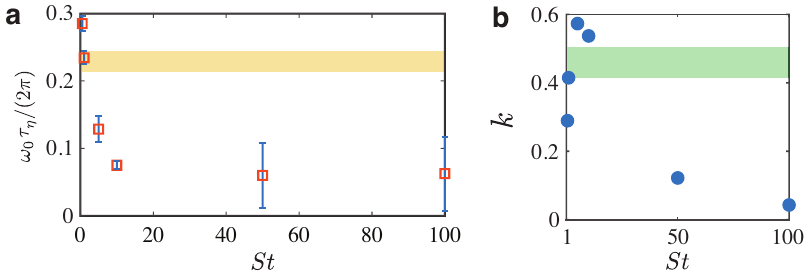}
	\caption{(a) The estimated $\omega_0$ (orange symbols) as a function of St. The error bars (blue bars) are drawn by estimating the standard deviation of $\omega_0$ detected as $a_e$ changes (within the range of $a_e<5$). The light orange shaded area highlights the range of $\omega_0$ detected under the condition of $\beta=3$ (light particle), St $=1$ (most prominent resonant), and Re$_{L} \in [25, 133]$. 
		(b) The parameter $k$ (symbols) from the superposition model as a function of St. The green shaded area highlights the range of $k$ detected under the condition of $\beta=3$ (light particle), St $=1$ (most prominent resonant), and Re$_{L} \in [25, 133]$.
	}
	\label{fig8}
\end{figure}
It has been reported that the preferential accumulation of light particles (or bubbles) clustering in the high vorticity region is pronounced under the condition of St $\approx1$ \cite{calzavarini2008dimensionalitysi, mathai2020bubblysi}, thus creating a perfect state of localization when there is no external force applied. This marks the starting point for achieving the final goal of turbulence manipulation, as the candidate particles need to first possess the ability to dive into the vortex before we enforce the desired behavior by adding external forces. One may wonder if the resonant phenomenon of light particles is still prominent when St deviates from 1. To address this question, we conducted multiple simulation runs varying St. The input parameters of the simulation are shown in the first row of Table.~\ref{simulation_parameters_changest}. The original velocity oscillation amplitude $v_0$ is depicted in Fig.~\ref{fig5}, while the corresponding enstrophy and normalized plots are presented in Fig.~\ref{fig6} and Fig.~\ref{fig7}, respectively.

When $ a \leq 10 $, we can still distinguish between the trapped (high enstrophy level) and escaped states (low enstrophy level) of particles, as depicted in Fig. \ref{fig6}(a-d). The natural frequency, $\omega_0$, can only be detected if the particle remains trapped in the vortex. However, when $ a > 10 $, the ensemble average enstrophy for particles remains low for various external forces, indicating the particles' free movement under the influence of the external force, as illustrated in Fig. \ref{fig6}(e-f). As shown in Fig. \ref{fig7}, for a given $a_e$, increasing $\omega$ gradually leads to a resonant phenomenon occurring for St $\leq 10$. Conversely, for St $>10$, the response parameter $v_0/v_{\text{free}}$ remains almost unchanged as $\omega$ and $a_e$ vary. We can still utilize the superposition model to fit the simulation data following the procedures outlined in Section C. The estimated $\omega_0$ as a function of St is presented in Fig. \ref{fig8}(a) (orange symbols). The error bars (blue bars) are drawn by estimating the standard deviation of $\omega_0$ detected as $a_e$ changes (within the range of $a_e<5$). The light orange shaded area in Fig. \ref{fig8}(a) highlights the range of $\omega_0$ detected under the condition of $\beta=3$ (light particle), St $=1$ (most prominent resonant), and Re$_{L} \in [25, 133]$. The parameter $k$ from the superposition model as a function of St is shown in Fig. \ref{fig8}(b), with the green shaded area highlighting the range of $k$ detected under the condition of $\beta=3$ (light particle), St $=1$ (most prominent resonant), and Re$_{L} \in [25, 133]$. When $k\rightarrow0$ (as St $>10$), we have $r\rightarrow0$, indicating $v_0^s -\rightarrow v_{\text{free}}$. It is worth noting that $v_{\text{free}} = a_e/\sqrt{\gamma_{\text{free}}^2+\omega^2}$ with $\gamma_{\text{free}} =1/\text{St}$, independent of $\omega_0$. This implies that $\omega_0$ can only be detected when $k$ well away from 0.

\section*{Section F: Changing $\beta$}

\begin{figure}
	\centering
	\includegraphics[width=\textwidth]{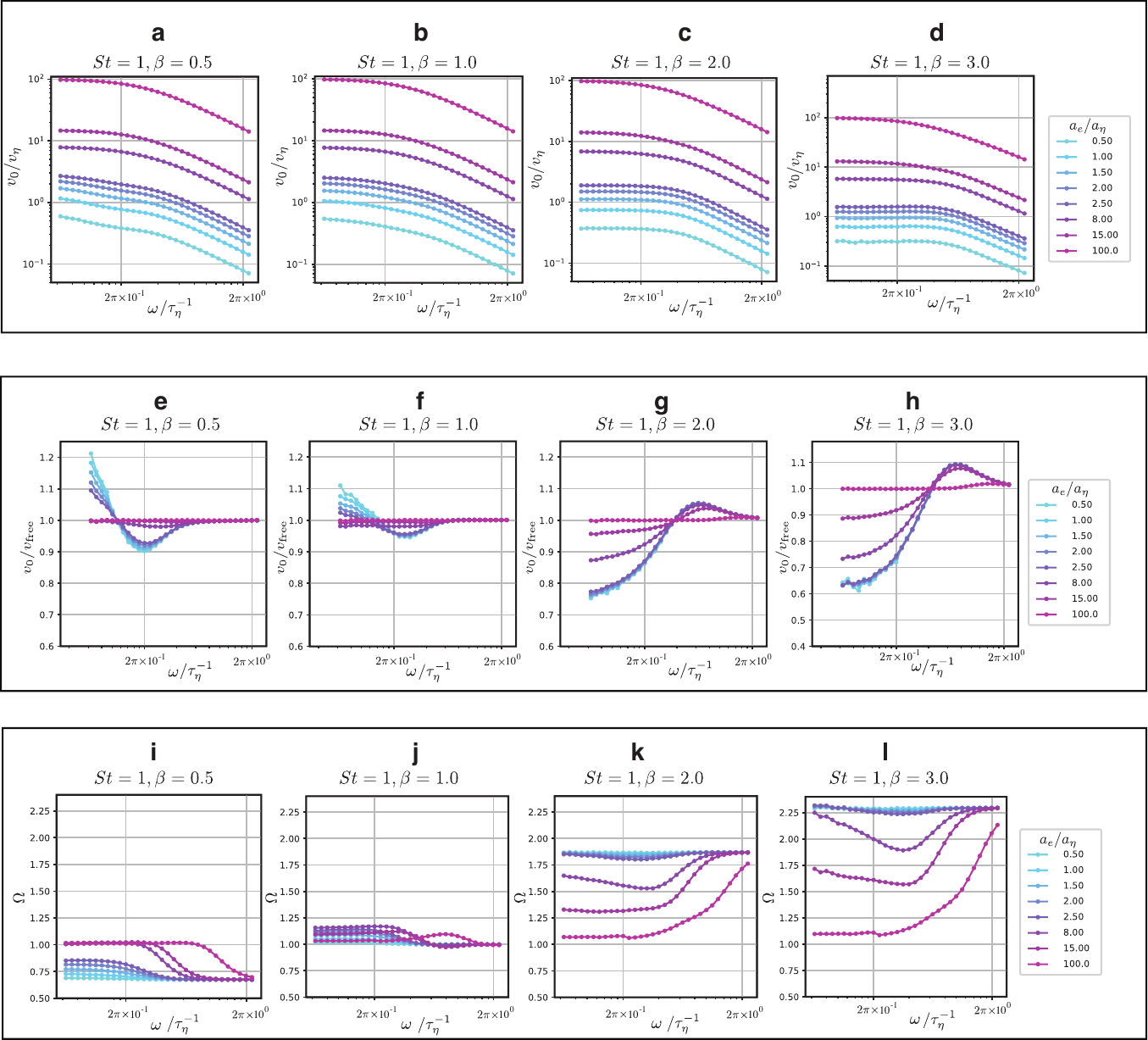}
	\caption{(a)-(d) The original velocity oscillation amplitude $v_0$ as a function of $a_e$ and $\omega$ for different values of $\beta$. (e)-(h) The normalized velocity oscillation amplitude $v_0/v_{\text{free}}$ as a function of $a_e$ and $\omega$. (i)-(l) Thee enstrophy level $\Omega$ as a function of $a_e$ and $\omega$. }
	\label{fig9}
\end{figure}

\begin{figure}
	\centering
	\includegraphics[width=\textwidth]{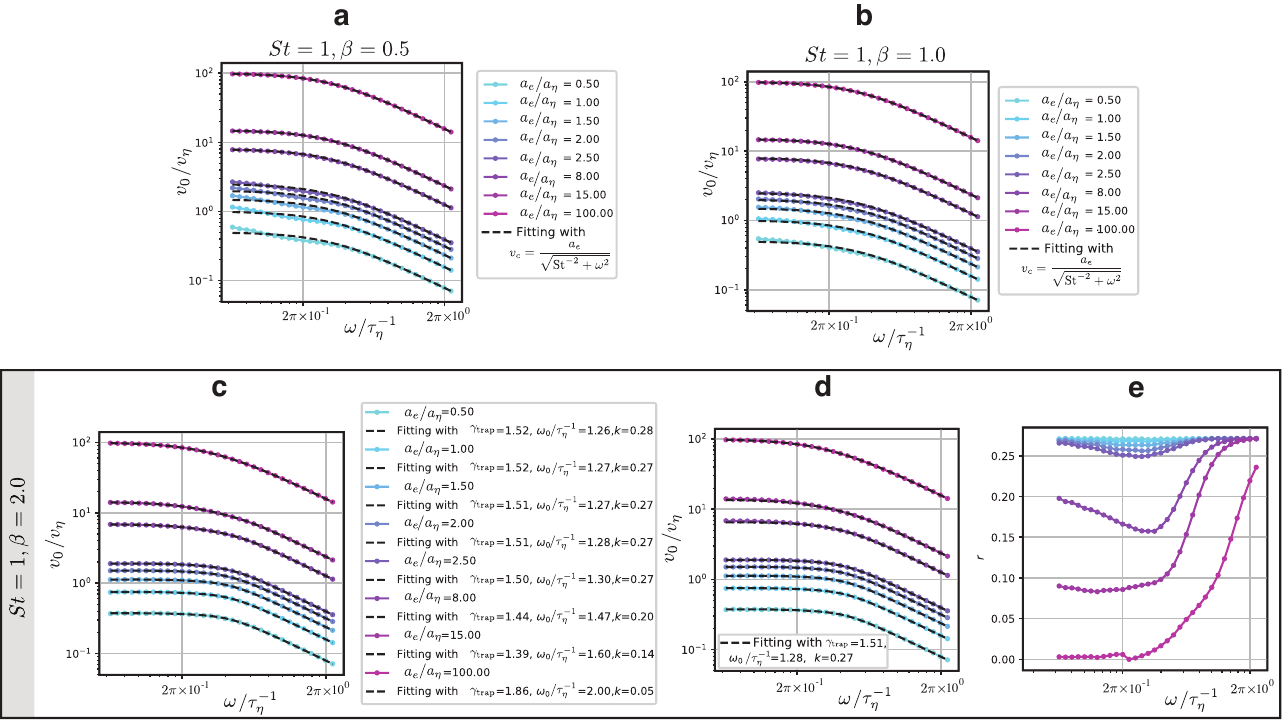}
	\caption{(a) Comaprison of theoretical prediction and simulation results for $\beta = 0.5$. (b) Comaprison of theoretical prediction and simulation results for $\beta = 1.0$. 
		The original velocity oscillation amplitude $v_0$ for $\beta=0.5$ and $\beta=1.0$ can be well captured by the conduction solution, $v_{\text{free}}$.
		(c)-(d) The superposition model results for $\beta=2.0$.
	}
	\label{fig10}
\end{figure}

In the main paper and the aforementioned discussions, $\beta$ is set to 3, representing light particles. However, the behavior differs significantly for neutral and heavy particles, which is beyond the scope of the current study. To provide a general overview, simulations are conducted with $\beta=0.5$ (heavy particle), $\beta=1.0$ (neutral particle), and $\beta=2.0$ (slightly light particle). The input parameters of the simulation are displayed in the first row of Table \ref{simulation_parameters_changebeta}.

The original velocity oscillation amplitude $v_0$ (Fig.~\ref{fig9}(a-d)), the normalized velocity oscillation amplitude $v_0/v_{\text{free}}$ (Fig.~\ref{fig9}(e-h)), and the enstrophy level $\Omega$ (Fig.~\ref{fig9}(i-l)) are presented. The trends observed for $\beta=2.0$ and $\beta=3.0$ are consistent, unlike those observed for $\beta=0.5$ and $\beta=1.0$. To identify the natural frequency, the value of $\beta$ can be extended beyond $\beta>2$, indicating a unique resonant phenomenon for slightly or extremely light particles, i.e., $\rho_p/\rho_f <4$. The original velocity oscillation amplitude $v_0$ for $\beta=0.5$ and $\beta=1.0$ can be well captured by the conduction solution, $v_{\text{free}}$, as shown in Fig.~\ref{fig10}(a-b). For $\beta=2.0$, following the superposition modelling procedures outlined in Section C, we can get the modelling results (see Fig.~\ref{fig10}(c-e)), which once again proves the robustness of the superposition model.

%% main parameter
\clearpage
\begin{table}\centering
\caption{The simulation parameters used in the main paper. Listed are the grid resolution $N^3$, cubic box size $L$, integral scale Reynolds number $\text{Re}_{L}$, Kolmogorov length $\eta$, grid spacing $dx$, kinematic viscosity $\nu$, energy dissipation $\epsilon $, Kolmogorov time $\tau_{\eta}$, temporal discretization $dt$, total simulation time $T$, the amplitude of the external force $a_e$, the angular frequency of the external force $\omega$, Stokes number of the particle St, the contrast between the density of the particle and that of the fluid $\beta$}
\begin{tabular}{lccccccccccccc}
	\centering
	$ N^3$ & $L$ & $\text{Re}_{L}$ & $\eta$ & $dx$ & $\nu$ & $\left< \epsilon \right>$ &$\tau_{\eta}$& $dt$ &$T$ & $a_e$ &$\omega$ & $\text{St}$ & $\beta$ \\
	\hline
	$256^3$ &  $2\pi$ & $58$ & $2.45\times 10^{-2}$ & $2.45\times 10^{-2}$ & $6.02\times 10^{-4}$ & $6.02\times 10^{-4}$ &$1$ &$5\times 10^{-3}$& $2000$ & $\left[1.0, 150.7 \right]$ & $[0.2, 7.0]$  & $1.0$ & $3.0$\\
	\hline
	\label{simulation_parameters_main}
\end{tabular}
\end{table}

%% change re parameter
\begin{table}\centering
\caption{The simulation parameters used in investigation of Reynolds number dependence.}
\begin{tabular}{lccccccccccccc}
	\centering
	$ N^3$ & $L$ & $\text{Re}_{L}$ & $\eta$ & $dx$ & $\nu$ & $ \epsilon $ &$\tau_{\eta}$& $dt$ &$T$ & $a_e$ &$\omega$& $\text{St}$ & $\beta$ \\
	\hline
	$128^3$ &  $2\pi$ & $25$ & $4.91\times 10^{-2}$ & $4.91\times 10^{-2}$ & $2.41\times 10^{-3}$ & $2.41\times 10^{-3}$ &$1$ &$1\times 10^{-2}$& $2500$ & $\left[0.5, 100 \right]$ & $[0.2, 7.0]$ & $1.0$ & $3.0$\\
	$512^3$ &  $2\pi$ & $133$ & $1.23\times 10^{-2}$ & $1.23\times 10^{-2}$ & $1.51\times 10^{-4}$ & $1.51\times 10^{-4}$ &$1$ &$25\times 10^{-3}$& $560$ & $\left[0.5, 100 \right]$ & $[0.2, 7.0]$ & $1.0$ & $3.0$\\
	\hline
	\label{simulation_parameters_changere}
\end{tabular}
\end{table}

%% change st parameter
\begin{table}\centering
\caption{The simulation parameters used in investigation of Stokes number dependence.}
\begin{tabular}{lccccccccccccc}
	\centering
	$ N^3$ & $L$ & $\text{Re}_{L}$ & $\eta$ & $dx$ & $\nu$ & $ \epsilon $ &$\tau_{\eta}$& $dt$ &$T$ & $a_e$ &$\omega$& St & $\beta$ \\
	\hline
	$256^3$ &  $2\pi$ & $58$ & $2.45\times 10^{-2}$ & $2.45\times 10^{-2}$ & $6.02\times 10^{-4}$ & $6.02\times 10^{-4}$ &$1$ &$5\times 10^{-3}$& $2000$ & $\left[0.5, 100 \right]$ & $[0.2, 7.0]$ & $0.5$ & $3.0$\\
	$256^3$ &  $2\pi$ & $58$ & $2.45\times 10^{-2}$ & $2.45\times 10^{-2}$ & $6.02\times 10^{-4}$ & $6.02\times 10^{-4}$ &$1$ &$5\times 10^{-3}$& $2000$ & $\left[0.5, 100 \right]$ & $[0.2, 7.0]$ & $1.0$ & $3.0$\\
	$256^3$ &  $2\pi$ & $58$ & $2.45\times 10^{-2}$ & $2.45\times 10^{-2}$ & $6.02\times 10^{-4}$ & $6.02\times 10^{-4}$ &$1$ &$5\times 10^{-3}$& $2000$ & $\left[0.5, 100 \right]$ & $[0.2, 7.0]$ & $5.0$ & $3.0$\\
	$256^3$ &  $2\pi$ & $58$ & $2.45\times 10^{-2}$ & $2.45\times 10^{-2}$ & $6.02\times 10^{-4}$ & $6.02\times 10^{-4}$ &$1$ &$5\times 10^{-3}$& $2000$ & $\left[0.5, 100 \right]$ & $[0.2, 7.0]$ & $10.0$ & $3.0$\\
	$256^3$ &  $2\pi$ & $58$ & $2.45\times 10^{-2}$ & $2.45\times 10^{-2}$ & $6.02\times 10^{-4}$ & $6.02\times 10^{-4}$ &$1$ &$5\times 10^{-3}$& $2000$ & $\left[0.5, 100 \right]$ & $[0.2, 7.0]$ & $50.0$ & $3.0$\\
	$256^3$ &  $2\pi$ & $58$ & $2.45\times 10^{-2}$ & $2.45\times 10^{-2}$ & $6.02\times 10^{-4}$ & $6.02\times 10^{-4}$ &$1$ &$5\times 10^{-3}$& $2000$ & $\left[0.5, 100 \right]$ & $[0.2, 7.0]$ & $100.0$ & $3.0$\\
	\hline
	\label{simulation_parameters_changest}
\end{tabular}
\end{table}

%% change beta parameter
\begin{table}\centering
\caption{The simulation parameters used in investigation of Stokes number dependence.}
\begin{tabular}{lccccccccccccc}
	\centering
	$ N^3$ & $L$ & $\text{Re}_{L}$ & $\eta$ & $dx$ & $\nu$ & $ \epsilon $ &$\tau_{\eta}$& $dt$ &$T$ & $a_e$ &$\omega$& St & $\beta$ \\
	\hline
	$256^3$ &  $2\pi$ & $58$ & $2.45\times 10^{-2}$ & $2.45\times 10^{-2}$ & $6.02\times 10^{-4}$ & $6.02\times 10^{-4}$ &$1$ &$5\times 10^{-3}$& $2000$ & $\left[0.5, 100 \right]$ & $[0.2, 7.0]$ & $1.0$ & $0.5$\\
	$256^3$ &  $2\pi$ & $58$ & $2.45\times 10^{-2}$ & $2.45\times 10^{-2}$ & $6.02\times 10^{-4}$ & $6.02\times 10^{-4}$ &$1$ &$5\times 10^{-3}$& $2000$ & $\left[0.5, 100 \right]$ & $[0.2, 7.0]$ & $1.0$ & $1.0$\\
	$256^3$ &  $2\pi$ & $58$ & $2.45\times 10^{-2}$ & $2.45\times 10^{-2}$ & $6.02\times 10^{-4}$ & $6.02\times 10^{-4}$ &$1$ &$5\times 10^{-3}$& $2000$ & $\left[0.5, 100 \right]$ & $[0.2, 7.0]$ & $1.0$ & $2.0$\\
	$256^3$ &  $2\pi$ & $58$ & $2.45\times 10^{-2}$ & $2.45\times 10^{-2}$ & $6.02\times 10^{-4}$ & $6.02\times 10^{-4}$ &$1$ &$5\times 10^{-3}$& $2000$ & $\left[0.5, 100 \right]$ & $[0.2, 7.0]$ & $1.0$ & $3.0$\\
	\hline
	\label{simulation_parameters_changebeta}
\end{tabular}
\end{table}

%\bibliography{mybib_new_f_nphys_2drbc}
%\end{document}

\end{widetext}

%\bibliographystyle{elsarticle-num}
%\bibliography{pnas-sample_si}

\end{document}